\documentclass[prb,twocolumn, showpacs,amsmath,amssymb,superscriptaddress,floatfix]{revtex4-1}
\usepackage{amsmath}
\usepackage{amssymb}
\usepackage{amsthm}
\usepackage{amsfonts}
\usepackage{physics}
\usepackage{dsfont}
\usepackage{xcolor}
\usepackage{listings}
\usepackage{enumerate}
\usepackage{latexsym}
\usepackage{mathdots}

\usepackage{psfrag}

\usepackage{color}

\usepackage{bm}
\usepackage{graphicx}
\usepackage{subfigure}

\newcommand{\beq}{\begin{equation}}
\newcommand{\eneq}{\end{equation}}

\newcommand{\anticomm}[2]{\left\{#1,#2\right\}}

\newcommand{\hmL}{\hat{\mathcal{L}}}
\newcommand{\cdh}{\hat{c}^\dagger}
\newcommand{\ch}{\hat{c}}
\newcommand{\ah}{\hat{a}}
\newcommand{\bh}{\hat{b}}
\newcommand{\bph}{\hat{b}'}

\newcommand{\rhoss}{\rho_{\texttt{ss}}}
\newcommand{\hamilss}{H_{\texttt{ss}}}

\newcommand{\mE}{\mathcal{E}}

\input{epsf}

\newcommand{\prodal}[2]{\underset{#1}{\overset{#2}{\prod}}}
\newcommand{\sumal}[2]{\underset{#1}{\overset{#2}{\sum}}}

\newcommand{\twopartdef}[2]
{
	\left\{
		\begin{array}{ll}
			#1 \\
			#2
		\end{array}
	\right.
}

\newcommand{\nn}{\nonumber}

\begin{document}

\tolerance 10000

\newcommand{\vk}{{\bf k}}


\title{An Extension of ETH to Non-Equilibrium Steady States}

\date{\today}
\author{Sanjay Moudgalya}
\affiliation{Department of Physics, Princeton University, Princeton, NJ 08544, USA}
\author{Trithep Devakul}
\affiliation{Department of Physics, Princeton University, Princeton, NJ 08544, USA}
\affiliation{Kavli Institute for Theoretical Physics, University of California, Santa Barbara CA 93106-4030, USA}
\author{D. P. Arovas}
\affiliation{Department of Physics, University of California at San Diego, La Jolla, CA 92093, USA}
\author{S. L. Sondhi}
\affiliation{Department of Physics, Princeton University, Princeton, NJ 08544, USA}

\begin{abstract}
We extend the notion of the Eigenstate Thermalization Hypothesis (ETH) to Open Quantum Systems governed by the Gorini-Kossakowski-Lindblad-Sudarshan (GKLS) Master Equation. 
We present evidence that the eigenstates of non-equilibrium steady state (NESS) density matrices obey a generalization of ETH in boundary-driven systems when the bulk Hamiltonian is non-integrable, just as eigenstates of Gibbs density matrices are conjectured to do in equilibrium. 
This generalized ETH, which we call NESS-ETH, can be used to obtain representative pure states that reproduce the expectation values of few-body operators in the NESS.
The density matrices of these representative pure states can be further interpreted as weak solutions of the GKLS Master Equation.
Additionally, we explore the validity and breakdown of NESS-ETH in the presence of symmetries, integrability and many-body localization in the bulk Hamiltonian. 
\end{abstract}

\maketitle
\section{Introduction}
An isolated ergodic or chaotic system tends, by definition to an equilibrium state at long times which is characterized by the values of the conserved quantities. However, an infinite system with precisely specified conserved quantities necessarily supports arbitrarily large subsystems where these conserved quantities fluctuate---as they can be exchanged with the the rest of the system. Equilibrium on the subsystems thus implies the existence of a (grand) canonical ensemble which yields the same equilibrium expectation values. 

For isolated quantum systems this argument leads us from a pure state description of the system to a mixed state ensemble description of the subsystem. Taking this to the extreme limit where the pure state is a single many body eigenstate, we arrive at the conjecture- known as the Eigenstate Thermalization Hypothesis (ETH) which most elegantly states that the reduced density matrix on small subsystems of an eigenstate is the Gibbs state.\cite{deutsch1991quantum, srednicki1994chaos, rigol2008thermalization, d2016quantum} From the properties of the Gibbs ensembles it then follows that the eigenstates must yield expectation values that depend on only on the value of intensive quantities such as energy density and number density in order for this identification to be consistent. But this logic can be now run in reverse. Let us say we know that a system is exactly described by the Gibbs state because it is entangled with a bath. Then ETH implies that we can replace the Gibbs state on the system by one of its strategically chosen eigenstates which yields the desired expectation values. So now we can go from the entangled system-bath description to a pure state description.\cite{khemani2014eigenstate} 

The question we address in this paper is whether a similar choice of description---between a density matrix and a pure state is possible for non-equilibrium steady states (NESS). The NESS we have in mind are density matrices $\rhoss$ describing transport obtained as stationary solutions to boundary driven Gorini-Kossakowski-Lindblad-Sudarshan (GKLS) master equations.\cite{gorini1976completely, lindblad1976generators} Implicitly, the system of interest is entangled with two families of baths which have been integrated out. There is more than one way to model the baths; for concreteness we have in mind that each family consists of infinitely many copies of identical baths that have interacted with the system in the so-called ``repeat interactions protocol''. \cite{bruneau2014repeated, grimmer2016open, strasberg2017quantum, pereira2018heat} The two families serve as sources and sinks of the transport currents. We show that for steady states in ergodic/chaotic systems one can trade the density matrix description for a pure state description of the system. These representative pure states reproduce the mixed state expectation values of  few body operators in the large system limit. Moreover they are {\it weak} solutions of the GKLS equation in the sense that their GKLS time evolutions yield time-independent expectation values of local operators in the infinite volume limit. For systems which are integrable in the bulk we also find weaker versions of some of these results.

In this work we build on two proximate developments. The first are the studies of boundary driven GKLS steady states by Prosen, \v{Z}nidari\v{c}, and co-workers which have led to large set of results on their existence and properties.\cite{prosen2009matrix, prosen2011open, vznidarivc2010thermalization, vznidarivc2015relaxation, vznidarivc2016diffusive, varma2017fractality} 
Notable among these are the identification of quadratic\cite{prosen2008third} and integrable GKLS systems\cite{ vznidarivc2010exact, vznidarivc2010matrix, vznidarivc2011solvable, prosen2011exact, prosen2013exterior, karevski2013exact, prosen2015matrix} and of long-range order and phase transitions out of equilibrium.\cite{prosen2008quantum, prosen2010long}
Closest to our point of departure is their demonstration in Ref.~[\onlinecite{prosen2013eigenvalue}] that the spectrum of $\rhoss$ exhibit Wigner-Dyson statistics when a closed form expression of $\rhoss$ is not available, which is believed to be the case when the bulk Hamiltonian is ergodic.\cite{karevski2013exact, prosen2015matrix} 
The second proximate development is the extension of the ETH idea to reduced density matrices in equilibrium, including ground states of systems in Ref.~[\onlinecite{khemani2014eigenstate}]. Specifically this paper argued that equilibrium reduced density matrices in the ground state and in excited states could be replaced by choosing ``representative states'' from among their eigenstates for the purposes of computing the expectation values of few body operators. In this work we extend this idea to NESS density matrices and we will refer to the overarching notion of trading density matrices for representative states as ``Non-Equilibrium Steady State Eigenstate Thermalization Hypothesis'' or NESS-ETH as an efficient mnemonic even though no thermalization is entailed outside the strictly equilibrium context.

This is perhaps also a good point to note that the literature on quantum NESS does not always take the GKLS equation as its point of departure---{\it prima facie} its simplfying assumptions do not have to hold in every quantum steady state. So the generality of the NESS-ETH we propose here for the full class of NESS requires investigation and we do not discuss this here.\cite{breuer2002f, jakvsic2002non, yukalov2011equilibration, weiss2012quantum, prosen2010exact, wang2017theory, yang2018nonequilibrium}
We also note recent work by Gullans and Huse \cite{gullans2018entanglement} which uses random circuits and thus some entirely new technology into the study of non-equilibrium transport.

%

%
The paper is organized as follows. 
In Sec.~\ref{sec:lindbladreview}, we review the GKLS Master Equation and the formalism to obtain the steady state density matrix for an open quantum system. 
In Sec.~\ref{sec:ethmotivation}, we introduce NESS-ETH in the eigenstates of $\rhoss$ as a natural generalization of ETH for the eigenstates of the Gibbs density matrix in equilibrium.
There we distinguish the existence of a representative pure state for operators with the statement of NESS-ETH in terms of the form of its matrix elements of local operators in the energy eigenbasis.
In Sec.~\ref{sec:testingeth}, we show evidence for the existence of a representative state as well as the ETH predicted form of the diagonal matrix elements.
There we discuss the properties of eigenstates of $\rhoss$, and show that the density matrices of the eigenstates are weak solutions to the GKLS Master Equation.  
In Sec.~\ref{sec:regimes}, we discuss the validity of NESS-ETH in the eigenstates of $\rhoss$ in the presence of symmetries, integrability, and many-body localization, where we show the apparent breakdown of NESS-ETH in the last two cases, up to subtleties we discuss.
We conclude in Sec.~\ref{sec:conclusions} with open questions. 
The appendices consist of technical details and a discussion on the construction of representative pure states in (integrable) quadratic open quantum systems.
\section{Review of the GKLS Master Equation}\label{sec:lindbladreview}
A typical setup of an open quantum system governed by the GKLS Master Equation involves two basic components: the \emph{system} $S$ (for example, a spin chain) and the \emph{environment} $E$.
%
%
The parts of the system that interact with the environment are referred to as the \emph{contacts} $C$, and the part that does not is the \emph{bulk} $B$.
The GKLS Master Equation governs the evolution of the bulk and the contacts (the system $S = B \cup C$) after tracing out the environment $E$.   
The equation reads\cite{lindblad1976generators, gorini1976completely, pearle2012simple}
\begin{equation}
    \dv{\rho}{t} = \hmL\left(\rho\right) \equiv \underbrace{-i\comm{H}{\rho}}_{\hmL_S\left(\rho\right)} + \underbrace{\sum_{\mu}{\left(2 L_\mu \rho L^\dagger_\mu - \anticomm{L^\dagger_\mu L_\mu}{\rho}\right)}}_{\hmL_C\left(\rho\right)},
\label{eq:lindbladeq}
\end{equation}
where $\rho$ is the density matrix, $H$ is the Hamiltonian of the system, and $\{L_\mu\}$ is the set of ``jump'' operators that couple the system $S$ to the environment $E$. 
We will refer to $\hmL$ in Eq.~(\ref{eq:lindbladeq}) as the Lindbladian superoperator which is meant to invoke the analogy to the Liouvillian superoperator in Hamiltonian systems which is denoted here by $\hmL_S$. As this terminology is not universally accepted in the literature, the reader is warned not to confuse $\hmL$ with the jump operators which are also frequently called the Lindblad operators. The superoperator defined by the jump operators $\hmL_C$ is often referred to as the dissipator and we will adopt that terminology as well.
In general the jump operators $\{L_\mu\}$ can have a support anywhere in $S$. 
However in this work, we only consider one-dimensional boundary-driven systems and the jump operators act only on the contacts $C_l$ and $C_r$ that live on the left and right ends of the chain, whereas the Hamiltonian acts on the entire system $S$. 
Thus the dissipator $\hmL_C$ in Eq.~(\ref{eq:lindbladeq}) can be split into two parts
\begin{equation}
    \hmL_C = \hmL_{C_l} + \hmL_{C_r}, 
\label{eq:LCleftright}
\end{equation}
where $\hmL_{C_l}$ (resp. $\hmL_{C_r}$) consist of the parts of $\hmL_C$ that involve jump operators that are supported on the left (resp. right) end of the chain. 
The non-equilibrium steady state (NESS) $\rhoss$ of the system is obtained using 
\begin{equation}
    \hmL\left(\rhoss\right) = 0,
\label{eq:lindbladeqness}
\end{equation}
which is presumably also the infinite-time density matrix of the system $S$ in the absence of purely imaginary eigenvalues of $\hmL$.\cite{albert2014symmetries}
The physical parameters of the environment $E$ such as the inverse temperature $\beta$ and the chemical potential $\mu$ are encoded in the jump operators $\{L_\mu\}$.
The simplest prescription for setting the temperature and chemical potential of the left and right contacts is obtained using the following intuition. \cite{vznidarivc2010thermalization}
When the contacts are decoupled from the rest of the system, the NESS has the following tensor product form: 
\begin{equation}
    \rhoss = \rho_{C_l} \otimes \rho_B \otimes \rho_{C_r}, 
\label{eq:nessdecoupled}
\end{equation}
where $\rho_{R}$ is the density matrix of the region $R$, $R \in \{C_l, B, C_r\}$.
Thus, when a given (left or the right) contact is set to temperature $\beta$, $\rho_{C_\zeta}$ ($\zeta \in \{l, r\}$) in Eq.~(\ref{eq:nessdecoupled}) should be Gibbs density matrix, i.e. $\left.\rho_G\right|_{C_\zeta}\left(\beta\right) = \exp(-\beta H_{C_\zeta})$, where $H_{C_\zeta}$ is the Hamiltonian restricted to the contact $C_\zeta$. 
In the case of a particle-number conserving system, when a  chemical potential $\mu$ is included in the bath, $\rho_{C_\zeta}$ is given by the generalized Gibbs density matrix $\left.\rho_{GG}\right|_{C_\zeta}\left(\beta, \mu\right) = \exp(-\beta \left(H_{C_\zeta} - \mu N_{C_\zeta}\right))$, where $N_{C_\zeta}$ is the number operator restricted to the contact $C_\zeta$.
We thus arrive at the condition\cite{vznidarivc2010thermalization}
\begin{equation}
    \hmL_{C_\zeta}\left(\rho\right) = 0 \iff \rho = \twopartdef{\left.\rho_G\right|_{C_\zeta}\left(\beta_\zeta\right)}{\left.\rho_{GG}\right|_{C_\zeta}\left(\beta_\zeta, \mu_\zeta \right)}, \;\;\; \zeta = l, r, 
\label{eq:reqdbaths}
\end{equation}
where $\beta_l, \mu_l$ and $\beta_r, \mu_r$ are the inverse temperatures and the chemical potentials of the left and right contacts respectively. 
Once Eq.~(\ref{eq:reqdbaths}) is satisfied for the contacts, the original problem can be viewed as the dynamics of an isolated system connected to two reservoirs (on the left and the right), each at their own temperature and chemical potentials.   
Note that the above prescription for setting the temperatures and chemical potentials of the leads is by no means unique. 
For example, another prescription involves examining the case when the jump operators are chosen symmetrically on the left and right ends of the chain.
One could obtain the required jump operators by imposing the condition that the steady state in such a system (or its reduced density matrix deep in the bulk) is a Gibbs reduced density matrix with the required parameters.  
While different prescriptions generically lead to different choices of jump operators, we do not expect a drastic change in the properties of the system, at least when the bulk Hamiltonian is ergodic.\cite{vznidarivc2010thermalization} 
\section{ETH and its extension}\label{sec:ethmotivation}
We briefly review ETH in isolated quantum system and motivate its extension to open quantum systems.
ETH was proposed as an explanation for thermalization in non-integrable isolated quantum systems.\cite{deutsch1991quantum, srednicki1994chaos, rigol2008thermalization, polkovnikov2011colloquium, d2016quantum}
In particular, the matrix elements of few-body operators in the energy eigenstates of a non-integrable model are conjectured to be of the form \cite{d2016quantum}
\begin{equation}
    \bra{m}\hat{O}\ket{n} = \bar{O}\left(E\right)\delta_{m, n} + R_{m, n} e^{-S\left(E\right)/2} f_O\left(E, \omega \right),
\label{eq:ethmatrixel}
\end{equation}
where $\hat{O}$ is a few-body operator, $\ket{m}$ and $\ket{n}$ are the energy eigenstates with energies $E_m$ and $E_n$, $E = \left(E_m + E_n\right)/2$, $\omega = E_m - E_n$, $R_{m,n}$ is a random variable with zero mean and unit variance, and $\bar{O}\left(E\right)$ is a smooth function of $E$.
In Eq.~(\ref{eq:ethmatrixel}), $S(E) \sim \log D$ for states in the middle of the spectrum, where $D$ is the Hilbert space dimension. 
Thus, the standard deviation of expectation values of operators in the eigenstates is expected to scale as $\sim 1/\sqrt{D}$ for eigenstates in the middle of the spectrum.\cite{beugeling2014finite}
A consequence of ETH is that the expectation value of an operator in a Gibbs state $Z^{-1}\exp\left(-\beta H\right)$ can be replaced by its expectation value in a particular eigenstate of $H$, which we call the representative pure state.\cite{khemani2014eigenstate} 
This is a consequence of the fact that
\begin{equation}
    Z^{-1} \textrm{Tr}\left(\hat{O} e^{-\beta H}\right) \sim Z^{-1}\int{\mathrm{d}E\ g(E) e^{-\beta E} \bar{O}\left(E\right)} \sim \bar{O}\left(E^\ast\right)
\label{eq:purereplace}
\end{equation}
where $g(E)$ is the density of states, and we have used that $\bar{O}\left(E\right)$ in Eq.~(\ref{eq:ethmatrixel}) is a smooth function of $E$ as well as the fact that $g(E) e^{-\beta E}$ peaks in a narrow energy window around $E = E^\ast$.
The latter is due to the fact that $g(E)$, by virtue of being the density of states of a local operator (here $\hamilss$), is of the form 
\begin{equation}
    g(E) \sim \exp\left(-\frac{\left(E - E_c\right)^2}{2\sigma^2}\right) \ .
\label{eq:dosform}
\end{equation}
Here $E_c$ is the energy of the center of the spectrum, and $\sigma^2$ scales with the bandwidth as a consequence of the central limit theorem, i.e.  $\sigma^2 \approx w N$, where $w$ is a constant. 
Thus $g(E) \exp(-\beta E)$ peaks at
\begin{equation}
    E^\ast\left(\beta\right) \approx E_c - \beta \sigma^2, 
\end{equation}
and the energy density at the peak is
\begin{equation}
    \epsilon^\ast\left(\beta\right) \equiv \frac{E^\ast}{N} \approx \epsilon_c - \beta w, 
\label{eq:repenergydensity}
\end{equation}
where $\epsilon_c$ is the energy density of the center of the spectrum.
Thus, the eigenstates of $H$ around energy density $\epsilon^\ast\left(\beta\right)$ are representative pure states of the Gibbs density matrix that reproduce expectation values of few body operators. 
An easy way to obtain the representative energy density $\epsilon^\ast$ is given by using the Hamiltonian $H$ as the few body operator. That is, 
\begin{equation}
    \epsilon^\ast\left(\beta\right) = \frac{Z^{-1} \textrm{Tr}\left(H e^{-\beta H}\right)}{N}.
\label{eq:repenergydensity2}
\end{equation}

We now motivate the extension of ETH to open quantum systems starting from the equilibrium case. In the following, we only consider Hamiltonians without particle number or spin conservation. The extension to particle number or spin preserving Hamiltonians is straightforward.
\subsection{Equilibrium}
When the temperatures of the left and right baths are the same ($\beta_l = \beta_r = \beta$), and the Hamiltonian on the system is non-integrable, we might expect that $\rhoss$ is the Gibbs state on the entire system,\cite{vznidarivc2010thermalization} i.e.
\begin{equation}
    \rhoss = \rho_G(\beta) =  \frac{1}{Z}\exp\left(-\beta H\right).
\label{eq:rhonessapprox}
\end{equation}
However, we note that Eq.~(\ref{eq:rhonessapprox}) cannot be exactly true, as explained in App.~\ref{app:validity}.
Despite this subtlety, we will refer to this steady state with no currents flowing through the system, as equilibrium. The meaning of this term will be more problematic when we allow the bulk Hamiltonian to be integrable or many body localized and we will discuss this in Section~\ref{sec:regimes}.
Nevertheless, it is reasonable to expect that $\rhoss$ in equilibrium is of the form
\begin{equation}
    \rhoss = e^{-\hamilss}, 
\label{eq:Hness}    
\end{equation}
where $\hamilss$ is a local Hamiltonian which agrees with $\beta H$ away from the boundaries, up to the constant $\log Z$. 
With this rewriting, $\rhoss$ has the form of a Gibbs density matrix with Hamiltonian $\hamilss$ and $\beta = 1$. With $\hamilss$ inheriting the non-integrability of $H$ we expect that it will exhibit ETH.
%
%
%
As a consequence expectation values of few-body operators in $\rhoss$ can be reproduced by the expectation value in a representative pure state, as illustrated in Eq.~(\ref{eq:purereplace}).
Calling the eigenvalues of $\hamilss$ the \emph{pseudoenergies}, 
we expect the pseudoenergy density of the representative state to be $\epsilon^\ast$ of Eq.~(\ref{eq:repenergydensity}) with $\beta = 1$. 
\subsection{Out of Equilibrium}
\begin{figure*}[ht!]
\centering
\begin{tabular}{cc}
\includegraphics[scale = 0.45]{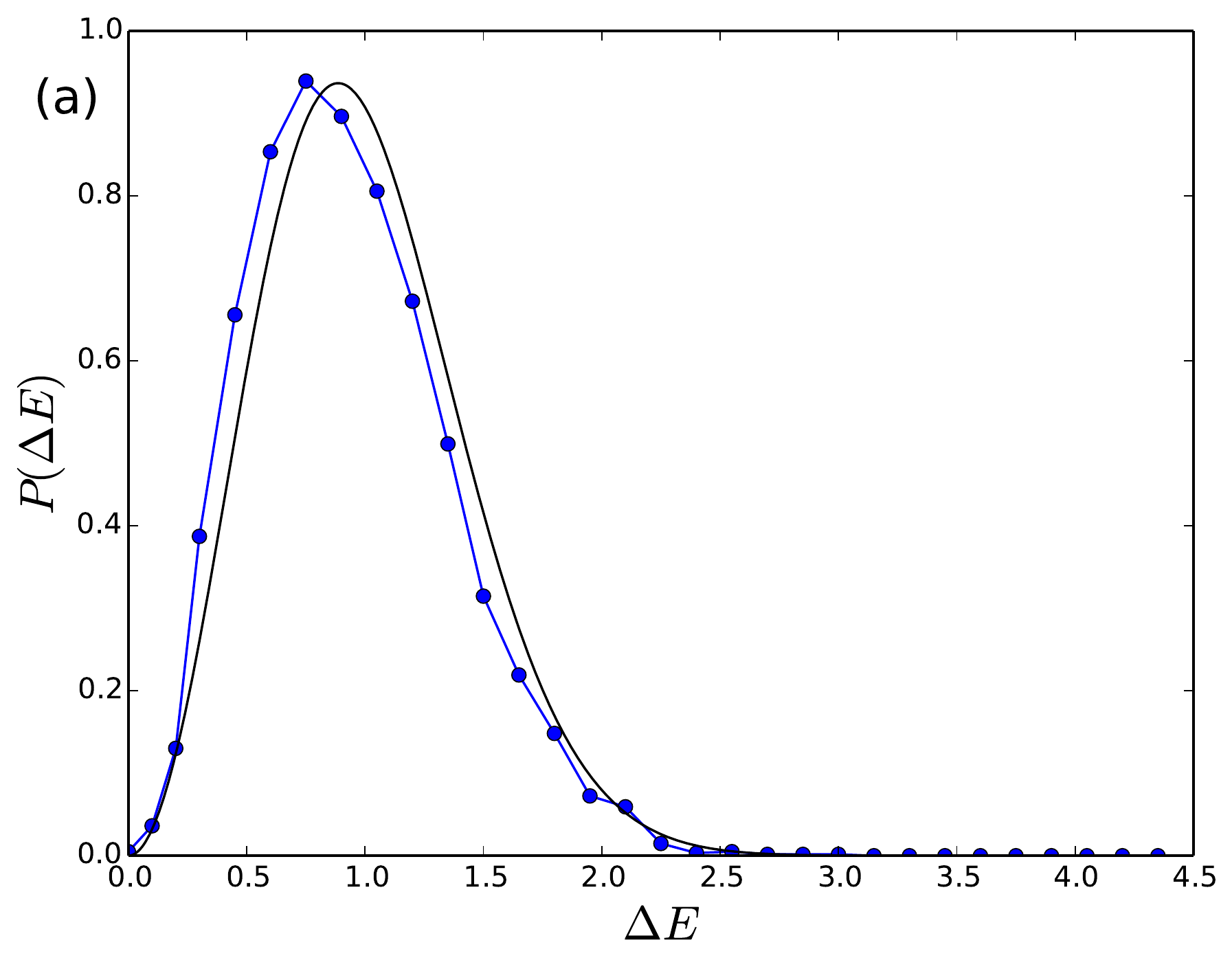}&\includegraphics[scale = 0.45]{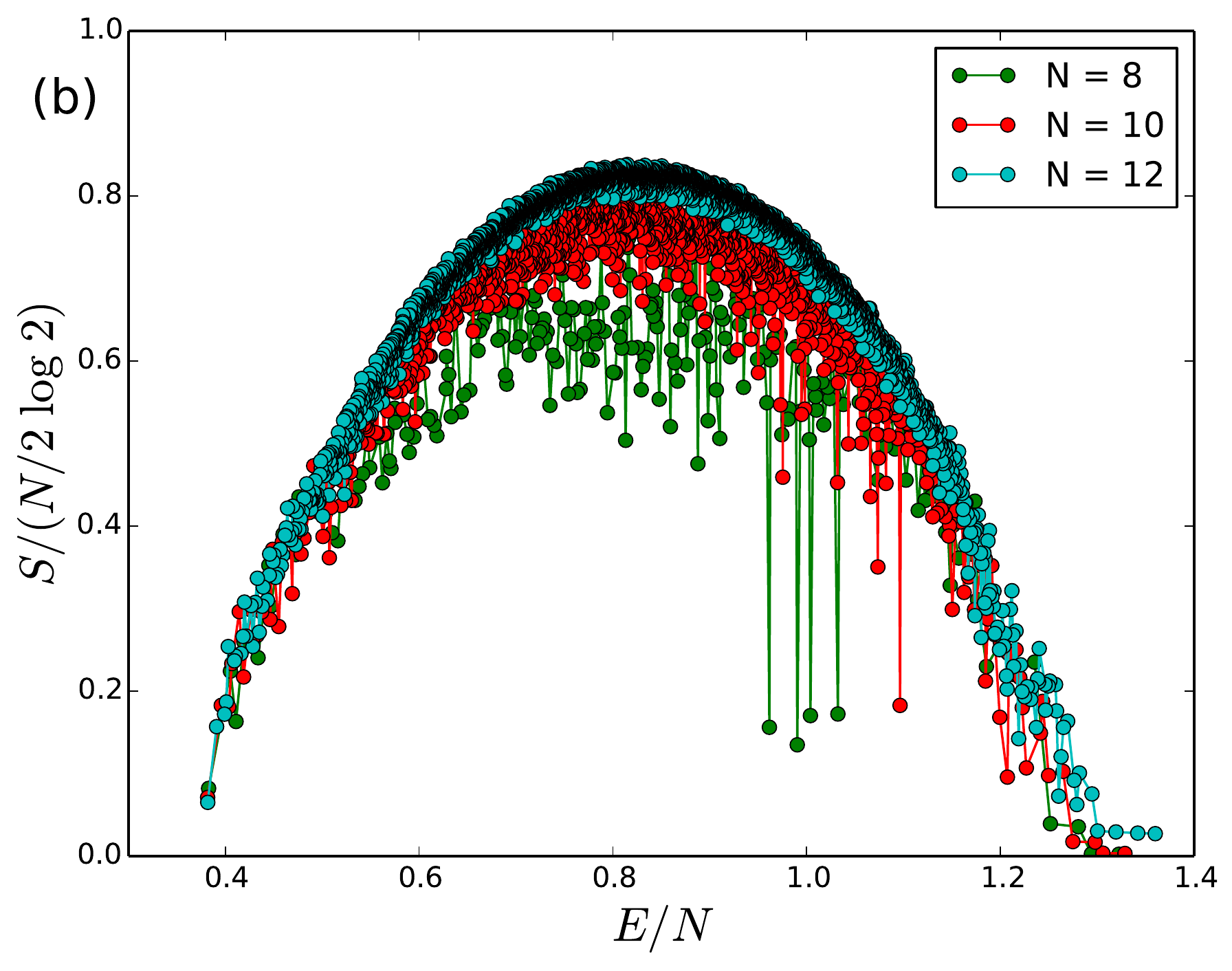}
\end{tabular}
\caption{(Color online) (a) Distribution $P(\Delta E)$, where $\Delta E = \{E_{n+1} - E_n\}$, where $\{E_n\}$ are the sorted pseudoenergies of $\rhoss$. Note that this follows the GUE distribution plotted in black. (b) Entanglement entropy of eigenstates of $\rhoss$. They scale as $~N$ for large systems, indicating a volume-law behavior.  
The parameters used for the plot are $(g, h, \beta_l, \beta_r, N) = (0.5, -1.05, 0.1, 1.0, 12)$.}
\label{fig:intlevelstats}
\end{figure*}
We now consider the situation away from equilibrium (i.e. $\beta_l \neq \beta_r$). 
Here, the steady state consists of heat (and particle/spin currents if particle number/spin is conserved) flowing from one end of the system to the other. 
If the bulk Hamiltonian is ergodic, for small $|\beta_l - \beta_r|$ (in the linear response regime), $\rhoss$ should be close to the form that encodes local equilibration \cite{mahajan2016entanglement, malouf2018information}
\begin{equation}
    \rhoss \approx \frac{1}{Z}\exp\left(-\sumal{i = 1}{N}{\beta_i H_i} - \lambda \frac{\beta_l - \beta_r}{2} J\right),
\label{eq:rhonessapproxnoneq}
\end{equation}
where $J$ is the heat current operator and $\beta_i$ is a ``local temperature" and $H_i$ is the piece of the Hamiltonian in the vicinity of site $i$ ($H = \sum_i{H_i}$). 
This form of $\rhoss$ cannot be exact for generic systems.\footnote{Although it is exactly true for certain systems.\cite{karevski2009quantum}} Among other things, it does not capture the long-range correlations that are expected to exist away from equilibrium.\cite{derrida2007non} Nevertheless we will draw inspiration from this form wherein both $H$ and $J$ in Eq.~(\ref{eq:rhonessapprox}) are sums of local operators, and conclude that we again expect $\rhoss$ to have the form of Eq.~(\ref{eq:Hness}), where $\hamilss$ is a local Hamiltonian. The reader may worry that we are ignoring the long-range correlations we just invoked but the latter are a subtle effect that decrease with system size\cite{derrida2007non} and we do not expect the Hamiltonian modifications needed to capture them materially affect the conclusions we reach in this paper.
%
%
%

Given this intuition, we expect ETH to hold for the eigenstates of $\hamilss$ (and thus those of $\rhoss$) even out of equilibrium. 
That is, we expect Eq.~(\ref{eq:ethmatrixel}) to hold for all few body operators, $\{E_m\}$ being the pseudoenergies of eigenstates $\{\ket{m}\}$ of $\rhoss$.  
The existence of ETH for $\hamilss$ away from equilibrium has a remarkable consequence: it enables a pure state description of the non-equilibrium current carrying steady state. 
More precisely the expectation values of few body operators in $\rhoss$ can be reproduced by their expectation values in an appropriate window of eigenstates of $\rhoss$, which are chosen by the peak in $g(E) \exp(-E)$ (see Eq.~(\ref{eq:purereplace})), where $E$ is the pseudoenergy of eigenstates of $\rhoss$ (energy of eigenstates of $\hamilss$) and $g(E)$ denotes its density of states.
\\

%
%
%

%
Before we move on to testing NESS-ETH, a few comments are in order. 
The strong version of NESS-ETH should imply the form of  Eq.~(\ref{eq:ethmatrixel}) for both diagonal and off-diagonal matrix elements of few body operators in the eigenstates of $\rhoss$ or equivalently $\hamilss$ at all locations in the spectrum except near its edges. Given NESS-ETH we can find a representative state by picking its pseudoenergy $\epsilon^\ast$ by setting $\beta = 1$ in Eqs.~(\ref{eq:repenergydensity}) and (\ref{eq:repenergydensity2}):
\begin{equation}
    \epsilon^\ast = \frac{\textrm{Tr}\left(\hamilss e^{-\hamilss}\right)}{N}.
\label{eq:reppseudodensity}
\end{equation}
However, for the purposes of obtaining a representative state, the eigenstates at other pseudoenergies, say in the middle of the spectrum, are not relevant as $\beta$ is not tunable---which is different from the equilibrium case. Consequently we generically do not expect a $1/\sqrt{D}$ scaling of the standard deviation of few body operators in the eigenstates with pseudoenergy $\epsilon^\ast$. 
Further, to obtain a representative state, we really only require that the fluctuations in operator expectation values between nearby eigenstates decrease with increasing $D$, which can in principle hold without the full set of eigenstates of $\rhoss$ satisfying the strong version of NESS-ETH. 
However, as we show in Sec.~\ref{sec:testingeth}, NESS-ETH (the form of the matrix elements of Eq.~(\ref{eq:ethmatrixel})) appears to hold for eigenstates of $\rhoss$, strengthening the case for existence of representative states.  
\section{Testing NESS-ETH}\label{sec:testingeth}
\begin{figure*}[ht!]
\centering
 \begin{tabular}{cc}
\includegraphics[scale = 0.45]{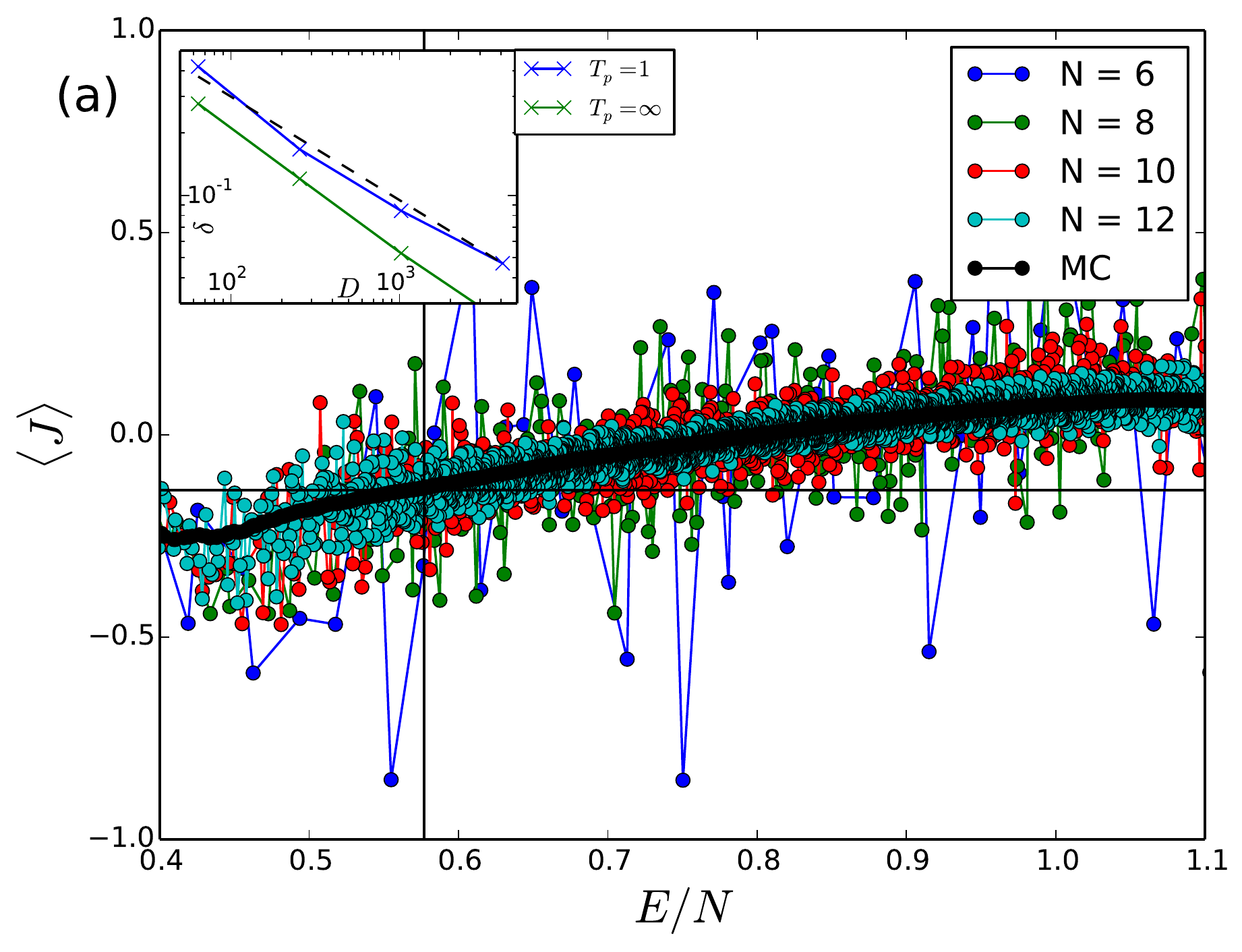}&\includegraphics[scale = 0.45]{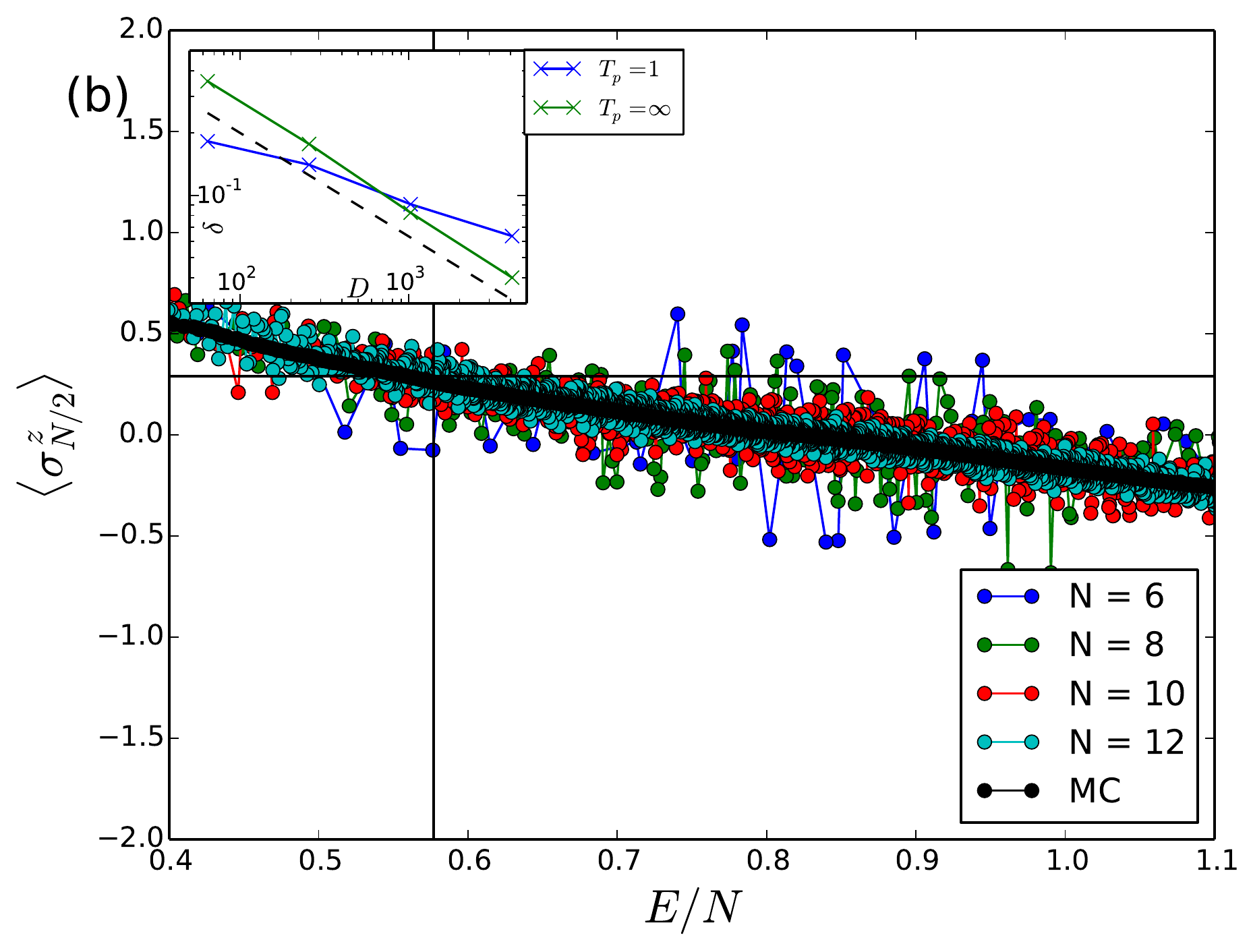}\\
\includegraphics[scale = 0.45]{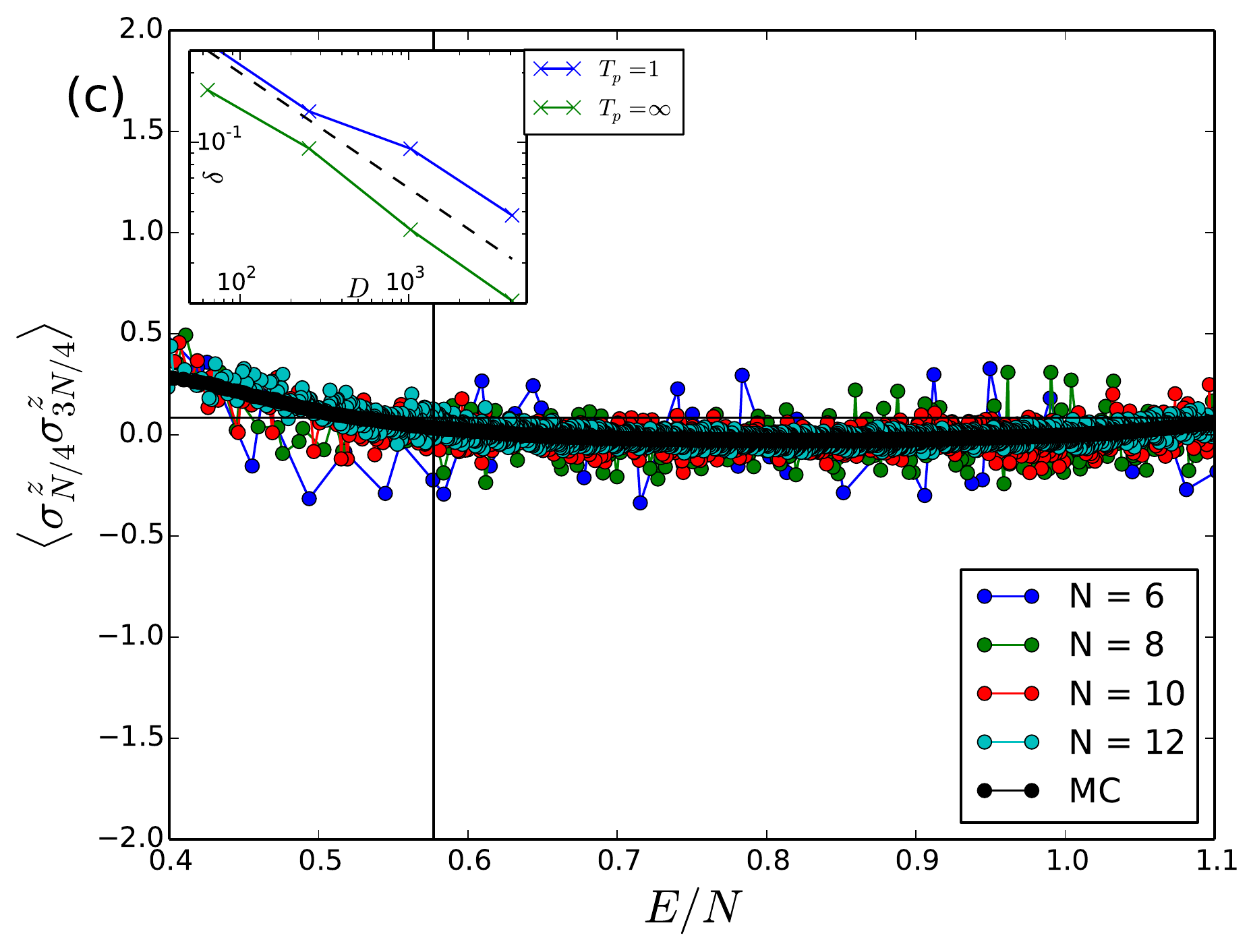}&\includegraphics[scale=0.45]{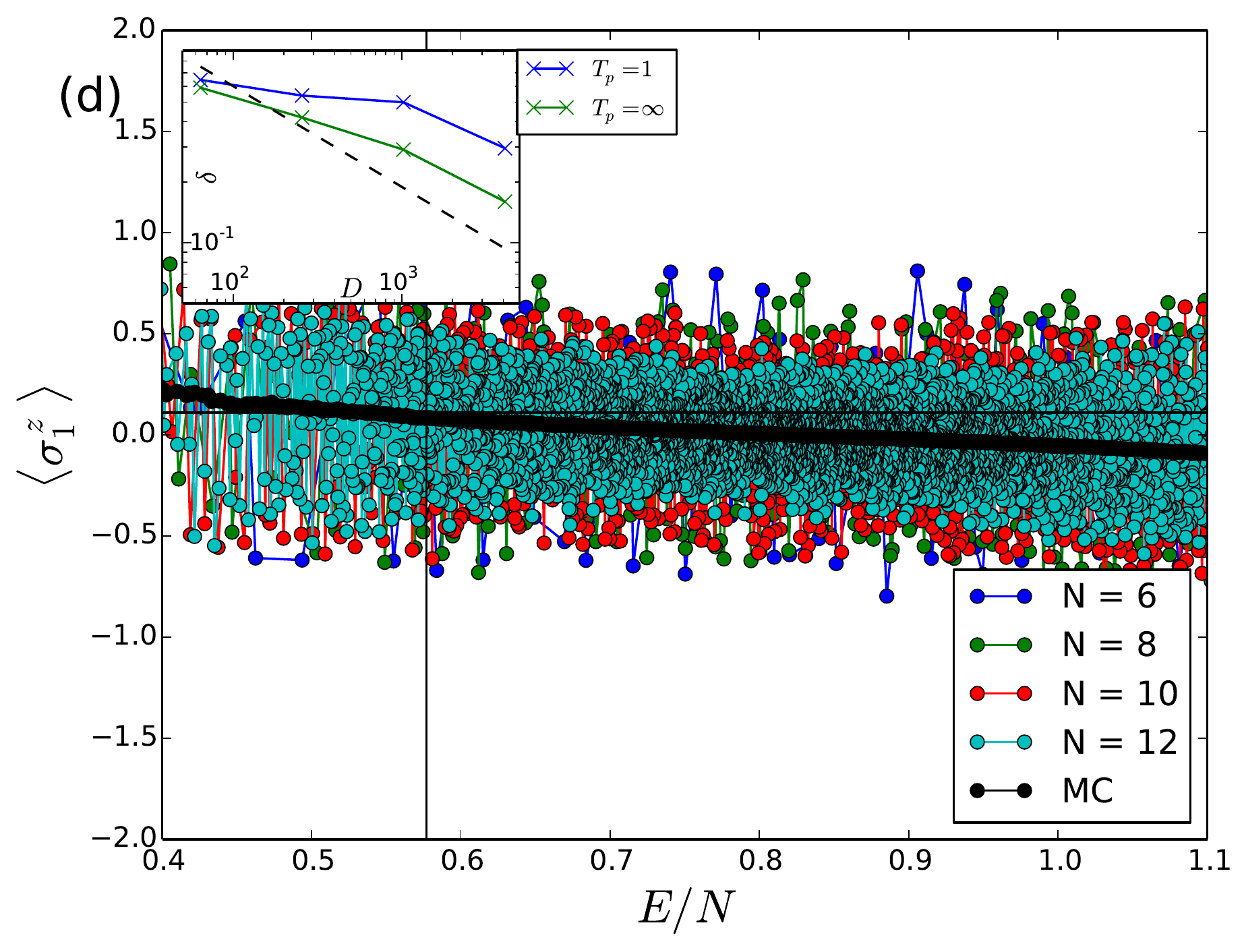}
\end{tabular}
\caption{(Color online) Expectation values and standard deviations of the operators (a) $J$  (b) $\sigma^z_{N/2}$ (c) $\sigma^z_{N/4}\sigma^z_{3N/4}$ (d) $\sigma^z_1$ in the eigenstates of $\hamilss$ as a function of pseudoenergy. (Main) The expectation values appear to converge to a smooth function of pseudoenergy with increasing system size, which is the ``microcanonical" (MC) value computed by averaging over an pseudoenergy window of $\Delta E = 0.025$. The expectation value in $\rhoss$ and the pseudoenergy density $\epsilon^\ast$ of the representative state are denoted by the horizontal and vertical lines respectively. (Inset) Standard deviations of the operators as a function of the Hilbert space dimension. The scaling of $1/\sqrt{D}$ is denoted by the black dashed line. The standard deviation appears to scales as $\sim 1/\sqrt{D}$ for eigenstates in the middle of the pseudoenergy spectrum ($T_p = \infty$), consistent with the predictions of NESS-ETH. The standard deviations around the eigenstates around the representative pseudoenergy ($T_p = 1$) also decay with increasing Hilbert space dimension, although they show a scaling slower than $1/\sqrt{D}$.
The data is shown for the parameters $(g, h, \beta_l, \beta_r) = (0.5, -1.05, 0.1, 1)$. }
\label{fig:inteth}
\end{figure*}
We now numerically test NESS-ETH away from equilibrium. In this paper we will only do so for the diagonal matrix elements. Since ETH is a statement about generic systems, we show our results for a fixed randomly picked set of parameters in Hamiltonian and the jump operators. 
\subsection{Model and Methods}\label{subsec:testingethmodels}
We first use a model with minimal symmetry, the boundary-driven tilted Ising model.\cite{prosen2009matrix}
The tilted Ising model of $N$ sites with open boundary conditions is given by the Hamiltonian
\begin{equation}
    H = \sumal{j = 1}{N-1}{\sigma^x_j \sigma^x_{j+1}} + \sumal{j = 1}{N}{\left(h \sigma^z_j + g\sigma^x_j\right)}, 
\label{eq:tiltedIsingmodel}
\end{equation}
which is known to be non-integrable for generic values of the fields $h$ and $g$.\cite{banuls2011strong}  
Since the model in Eq.~(\ref{eq:tiltedIsingmodel}) is inversion-symmetric, its energy levels show GOE statistics within each inversion symmetry sector. \cite{banuls2011strong, prosen2009matrix}
Furthermore, it does not conserve spin and thus does not support a spin-current.
The heat current operator for the tilted Ising model reads\cite{prosen2009matrix}
\begin{equation}
    J \equiv \frac{h}{N-2}\sumal{j = 2}{N-1}{J_j} = \frac{h}{N-2} \sumal{j = 2}{N - 1}{\sigma^y_{j} \left(\sigma^x_{j+1} - \sigma^x_{j-1}\right)}.
\label{eq:heatcurrent}
\end{equation}
To drive the system away from equilibrium, we introduce jump operators at the left and right ends that correspond to different temperatures $\beta_l$ and $\beta_r$, breaking inversion symmetry as a result.
For simplicity of numerical computations, we choose the left and right contacts to be composed of one physical site each. 
The jump operators $\{L_\mu\}$ for the required temperatures can be obtained using the condition of Eq.~(\ref{eq:reqdbaths}), and they read (see App.~\ref{app:jump})
\begin{eqnarray}
    &&L_1 = \sqrt{\Gamma^{(l)}_1}\tau^+_1 \;\;\; L_2 = \sqrt{\Gamma^{(l)}_2}\tau^{-}_1 \nn \\
    &&L_3 = \sqrt{\Gamma^{(r)}_1}\tau^+_N \;\;\; L_4 = \sqrt{\Gamma^{(r)}_2}\tau^-_N,
\label{eq:IntLindblad}
\end{eqnarray}
where
\begin{eqnarray}
    &\tau^\pm_j \equiv \frac{1}{2}\left(-\sigma^z_j\sin\theta  + (1 + \cos\theta) \sigma^\pm_j - (1 - \cos\theta)\sigma^\mp_j\right), \nn \\
    &\sin\theta = \frac{g}{\sqrt{g^2 + h^2}},\;\;\; \cos\theta = \frac{h}{\sqrt{g^2 + h^2}}.
\end{eqnarray}
With this definition of jump operators, the inverse temperatures read (see App.~\ref{app:jump})
\begin{equation}
    \beta_\zeta = -\frac{1}{2 \sqrt{h^2 + g^2}}\log\left(\frac{\Gamma^{(\zeta)}_1}{\Gamma^{(\zeta)}_2}\right),\;\; \zeta = l, r. 
\label{eq:tiltedtemp}
\end{equation}
Even though the inverse temperature $\beta_\zeta$ in Eq.~(\ref{eq:tiltedtemp}) only depends on the \emph{ratio} between $\Gamma^{(\zeta)}_1$ and $\Gamma^{(\zeta)}_2$, the GKLS master equation and hence $\rhoss$ depends on the strength of each of $\Gamma^{(\zeta)}_1$ and $\Gamma^{(\zeta)}_2$. 
Indeed, setting the strengths of the jump operators $\{L_\mu\}$ to be very large compared to the strength of the Hamiltonian $H$ results in a highly degenerate manifold of approximate steady states (since $\rhoss$ is then only constrained by the boundary jump operators), which result in strong finite size effects.
Throughout this work, we thus choose $\Gamma^{(l)}_2 = \Gamma^{(r)}_2 = 1$ and use $\Gamma^{(l)}_1$ and $\Gamma^{(r)}_1$ to tune $\beta_l$ and $\beta_r$. 
To obtain $\rhoss$ for the system we are studying, we exploit the fact that $\rhoss$ has an efficient Matrix Product Operator (MPO) representation. \cite{prosen2009matrix, mahajan2016entanglement} 
While several methods have been developed to study open quantum systems with tensor networks,\cite{prosen2009matrix, bonnes2014superoperators, cui2015variational} we use a variant of the Time Evolution Block Decimation (TEBD) algorithm\cite{vidal2004efficient, daley2004time, verstraete2008matrix} starting with a random initial density matrix $\rho$ to obtain the Matrix Product Operator (MPO) Representation of $\rhoss$. 
Details of the numerics are given in App.~\ref{app:tebddetails}.
We then obtain the $2^N$-dimensional matrix representation of $\rhoss$, and diagonalize it to study its eigenstates. 
This allows us to study the eigenstates of $\rhoss$ for much larger system sizes (easily up to $N = 12$ without any symmetries) as opposed to naive exact diagonalization of the Lindbladian. 
\footnote{Note that the $\rhoss$ obtained using TEBD necessarily has a Trotter error.\cite{vidal2004efficient} This results in a mixing of the eigenstates of the true $\rhoss$ especially in the middle of its spectrum where the gaps between the eigenvalues of $\rhoss$ is exponentially small in $L$. However, such a mixing can be made arbitrarily small by decreasing the TEBD time step (resulting in a slower convergence), and we have checked that it does not affect any of the results in this paper. }
\subsection{Level statistics and Entanglement entropy}\label{sec:testingethdiag}
We first test for level repulsion in the eigenstates of $\hamilss$. 
As shown in Fig.~\ref{fig:intlevelstats}a, we find that away from equilibrium, the pseudoenergy level spacings show Gaussian Unitary Ensemble (GUE) statistics,\cite{atas2013distribution} as a consequence of the broken time-reversal symmetry away from equilibrium.
Indeed, if $\rhoss$ has the form of Eq.~(\ref{eq:rhonessapproxnoneq}), the current operator breaks time-reversal symmetry.
The connection of the level statistics of $\rhoss$ to the integrability of the NESS was explored for several models in Ref.~[\onlinecite{prosen2013eigenvalue}]. 
There it was conjectured that the level statistics of $\rhoss$ follows GUE in generic cases when $\rhoss$ does not have a closed-form expression and exhibits Poisson level statistics in integrable cases.
Random matrix level statistics are necessary but not sufficient for ETH to hold. \footnote{See Ref.~[\onlinecite{chang2018evolution}] for such an example in the context of reduced density matrices of isolated quantum systems.}
Hence the observation of the level statistics predicted by random matrix theory is suggestive but not dispositive in our search for an NESS-ETH for the eigenstates of $\rhoss$. 
Further support comes from the entanglement entropy of the eigenstates of $\rhoss$ shown in Fig.~\ref{fig:intlevelstats}b which exhibits a \emph{volume-law} similar to ones observed in typical isolated non-integrable systems.\cite{nakagawa2018universality}
\subsection{Expectation values of operators}
We now test the expectation values of few-body operators in the eigenstates of $\rhoss$. 
We plot the expectation value and its standard deviation of the current operator $J$ (Fig.~\ref{fig:inteth}a), an operator in the middle of the system $\sigma^z_{N/2}$ (Fig.~\ref{fig:inteth}b),  a non-local two-site operators in the middle of the system $\sigma^z_{N/4} \sigma^z_{3N/4}$ (Fig.~\ref{fig:inteth}c), and an operator in the contacts $\sigma^z_1$ (Fig.~\ref{fig:inteth}d). 
The insets in Fig.~\ref{fig:inteth} show the scaling of the standard deviation with the Hilbert space dimension $D$ for pseudoenergy densities specified by pseudotemperatures $\beta_p = 1$ ($T_p = 1$) and $\beta_p = 0$ ($T = \infty$). As explained in the previous section, the decay of the former is important for the existence of a representative pure state and the $1/\sqrt{D}$ scaling of the latter is important for NESS-ETH of the form of Eq.~(\ref{eq:ethmatrixel}) to hold for eigenstates of $\rhoss$. 
As evident in Fig.~\ref{fig:inteth}, we find evidence of that the diagonal matrix elements of all these operators are of the form of Eq.~(\ref{eq:ethmatrixel}), along with the existence of representative pure states. 
However, $\sigma^z_1$ appears to show a larger standard deviation and stronger finite-size effects as seen in Fig.~\ref{fig:inteth}d, although the scaling of the standard deviation appears to be $1/\sqrt{D}$, consistent with the predictions of NESS-ETH. 
We have checked that only operators that have supports in the contacts appear to show the larger standard deviation of the form in Fig.~\ref{fig:inteth}d, and not for operators that are in the bulk but close to the contacts (for example $\sigma^z_2$ in the present case). 
We now focus on the representative pure state that reproduces local properties of $\rhoss$. 
To verify that these indeed represent $\rhoss$, we test certain physical properties of the eigenstates. %
In particular, if the eigenstates carry current, the current through one part of the bulk should be the same as the current through another part, as is exactly true in $\rhoss$:
\begin{equation}
    \textrm{Tr}\left(\rhoss J\right) = \textrm{Tr}\left(\rhoss J_{\frac{N}{2}}\right).
\label{eq:currentcutness}
\end{equation}
\begin{figure*}[ht!]
\centering
 \begin{tabular}{cc}
\includegraphics[scale = 0.45]{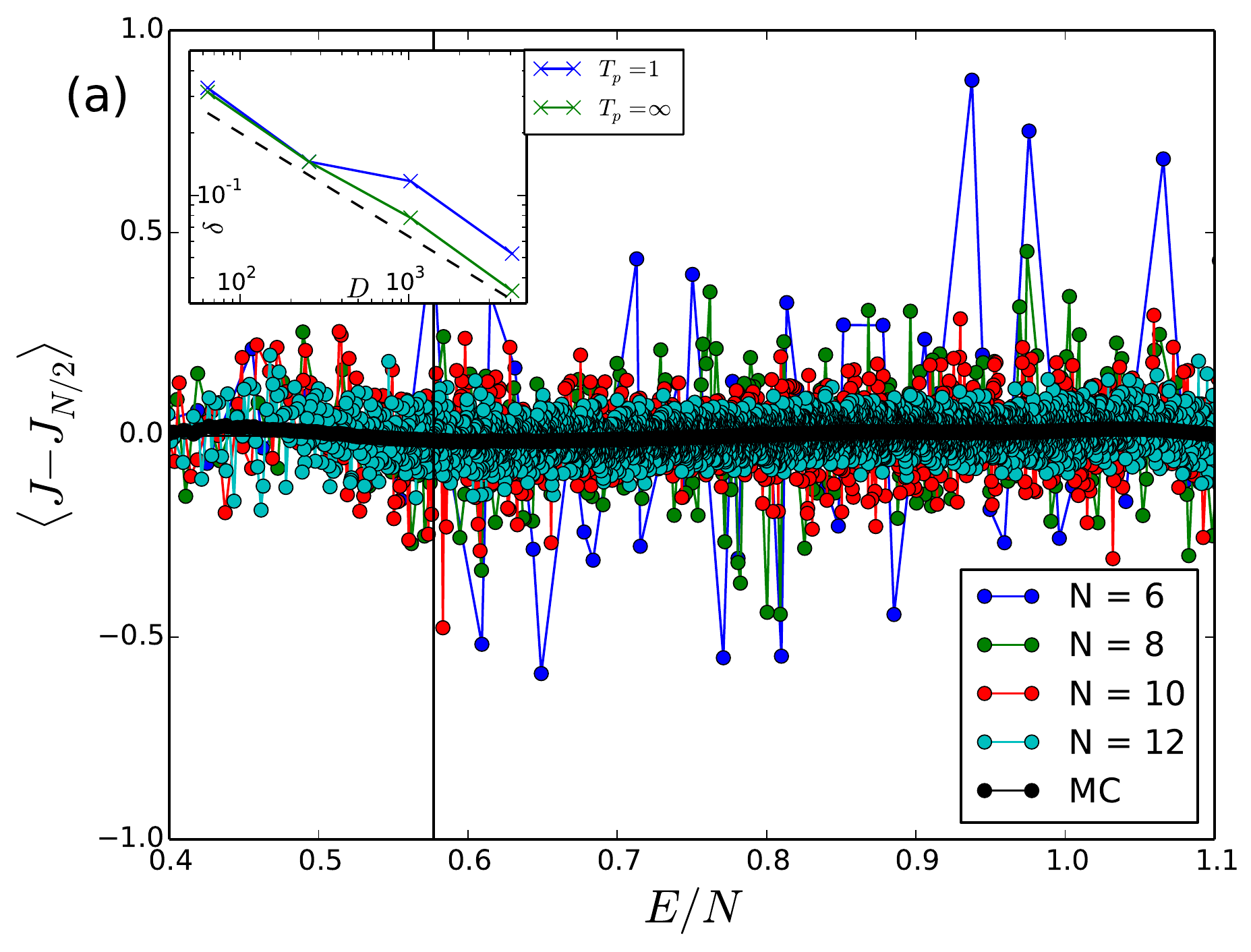}&\includegraphics[scale = 0.45]{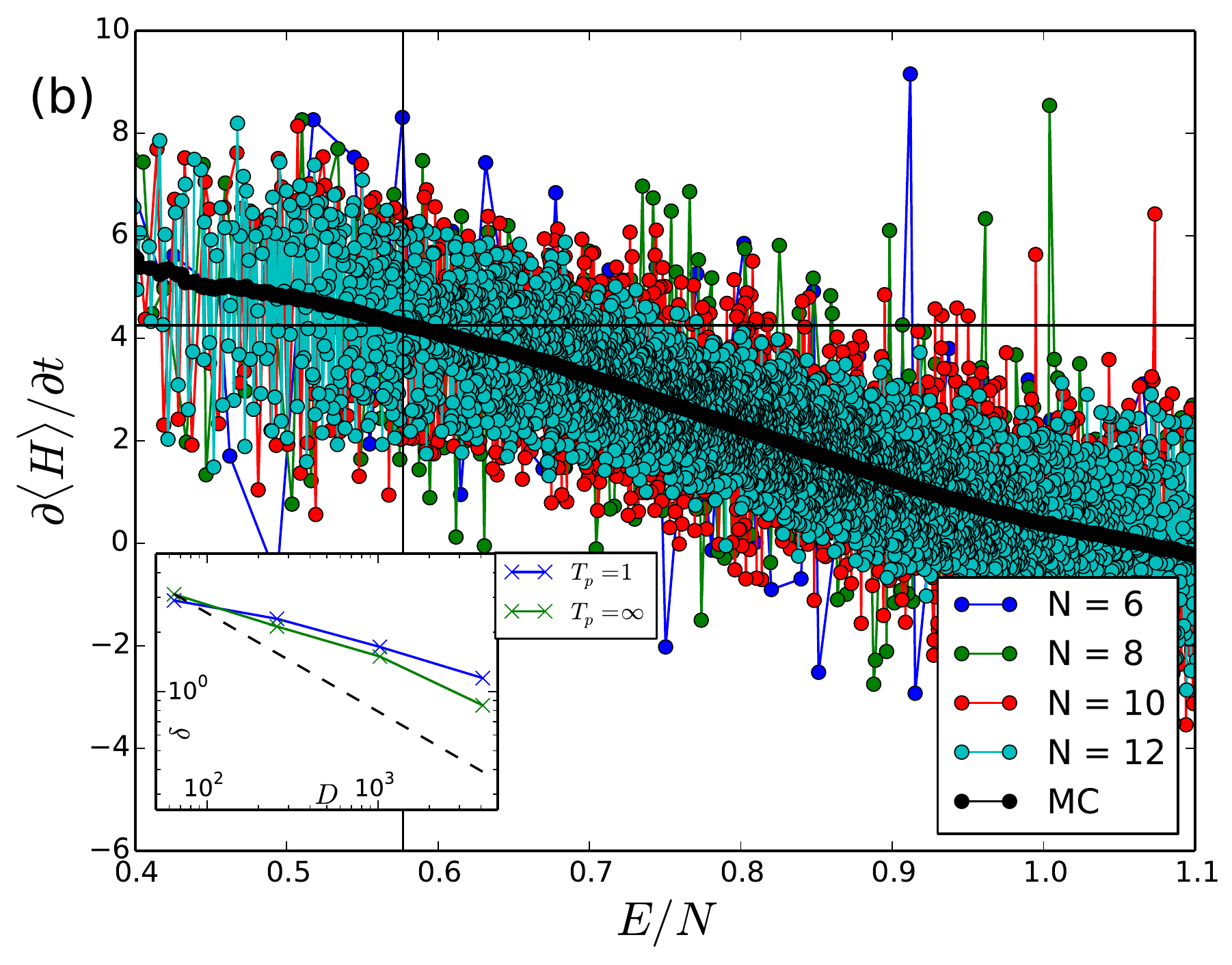}
\end{tabular}
\caption{(Color online) (a) $\langle J - J_{N/2} \rangle$ and (inset) its standard deviation in the eigenstates of $\rhoss$. (b) $\dv{\langle H \rangle}{t}$ for $\langle H \rangle$ computed in the eigenstates of $\rhoss$. Its vanishing at the representative pseudoenergy shows that $\rho_n = \ket{n}\bra{n}$ is a weak solution of the GKLS Master Equation, where $\ket{n}$ is the representative eigenstate. See caption of Fig.~\ref{fig:inteth} for details on the labels and parameters used.}
\label{fig:currenteig}
\end{figure*}
Thus, in the representative pure states, we expect that
\begin{equation}
    \bra{n}J\ket{n}\approx \bra{n}J_{\frac{N}{2}}\ket{n} \approx \textrm{Tr}\left(\rhoss J_{\frac{N}{2}}\right),
\label{eq:currenteigenstate}
\end{equation}
with Eq.~(\ref{eq:currenteigenstate}) being exactly true as $N \rightarrow \infty$.
Fig.~\ref{fig:currenteig}a plots the expectation values and the standard deviation of $\left(J - J_{N/2}\right)$ in the eigenstates of $\rhoss$, which we expect to be close to $0$ for states around pseudoenergy density $\epsilon^\ast$. 
As seen in Fig.~\ref{fig:currenteig}a, we indeed find that this is true for all eigenstates of $\rhoss$ (not just the representative pure state). 
Thus, all the eigenstates of $\hamilss$ (and thus $\rhoss$) carry a current. 
Note that the eigenstates of $\hamilss$ can carry a current even though $\hamilss$ has open boundary conditions because $J$ is the current operator of $H$, not of $\hamilss$. 
\subsection{Weak solution of the GKLS equation}
Since we have obtained a pure state that describes the local properties of $\rhoss$, one might wonder if the representative state (density matrix) is in some sense a solution of the GKLS Master Equation.
Of course, this cannot be true for any finite system size $N$ because of theorems that ensure the uniqueness of $\rhoss$.
However, the spectral gap $\Delta$ of the Lindbladian $\hmL$ above the steady state vanishes as $1/N^\delta$ for some $\delta > 0$ in all boundary-driven systems\cite{vznidarivc2015relaxation} which potentially allows the existence of multiple asymptotically exact steady state solutions in the large $N$ limit. Such solutions would presumably be \emph{strong} solutions in the sense that they would involve a specification of density matrices satisfying the GKLS equation.
%

We are not able to make sense of this idea at this time. Instead, here we show the existence of what we term  \emph{weak} steady state solution by studying the time-dependence of expectation values of local operators which are a special case of few body operators; we expect the result is true for the larger class.
That is, we compute 
\begin{equation}
    \dv{\langle O \rangle}{t} = \dv{}{t}\left(\textrm{Tr}\left(\rho O\right)\right) = \textrm{Tr}\left(\dv{\rho}{t}O\right).
\label{eq:expectevol}
\end{equation}
Expectation values of operators in the steady state are by definition time-independent, which can be seen by setting $\rho = \rhoss$ in Eq.~(\ref{eq:expectevol}) (and using $\mathrm{d}\rho_{ss}/dt = 0$). 
We can then probe the expectation value of operators in eigenstates of $\rhoss$ by setting $\rho = \rho_n \equiv \ket{n}\bra{n}$ in Eq.~(\ref{eq:expectevol}).
Using Eqs.~(\ref{eq:expectevol}) and (\ref{eq:lindbladeq}) we obtain
\begin{equation}
    \textrm{Tr}\left(\dv{\rho_n}{t}O\right) = \bra{n}\left(i [H, O] + 2 L_\mu O L_\mu^\dagger - \{L_\mu^\dagger L_\mu, O\}\right)\ket{n}.
\label{eq:expecteigsol}
\end{equation}
If $\ket{n}$ is indeed a representative pure state of $\rhoss$, we expect Eq.~(\ref{eq:expecteigsol}) to vanish.
In Fig.~\ref{fig:currenteig}b, we plot the quantity of Eq.~(\ref{eq:expecteigsol}) with $O = H$ for all the eigenstates $\ket{n}$ of $\rhoss$.
Note that it vanishes for $\ket{n}$ around the pseudoenergy density $\epsilon^\ast$, showing that $\rho_n$ is a \emph{weak} steady state solution of the GKLS Master equation, i.e. expectation values of local operators in $\rho_n$ are time-independent in the limit $N \rightarrow \infty$.
\section{Regimes of Validity}\label{sec:regimes}
\begin{figure*}[ht!]
\centering
 \begin{tabular}{ccc}
\includegraphics[scale = 0.45]{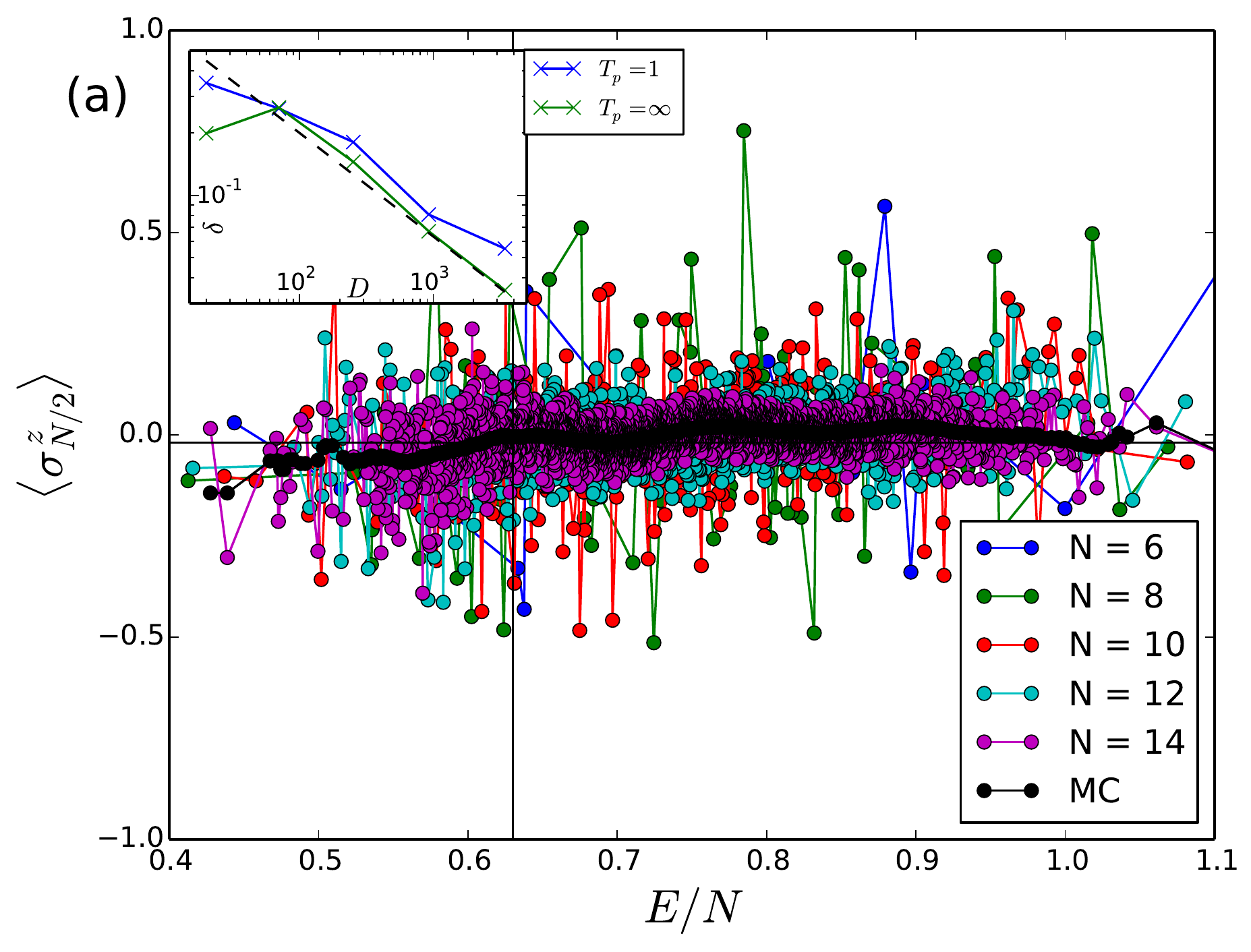}&\includegraphics[scale = 0.45]{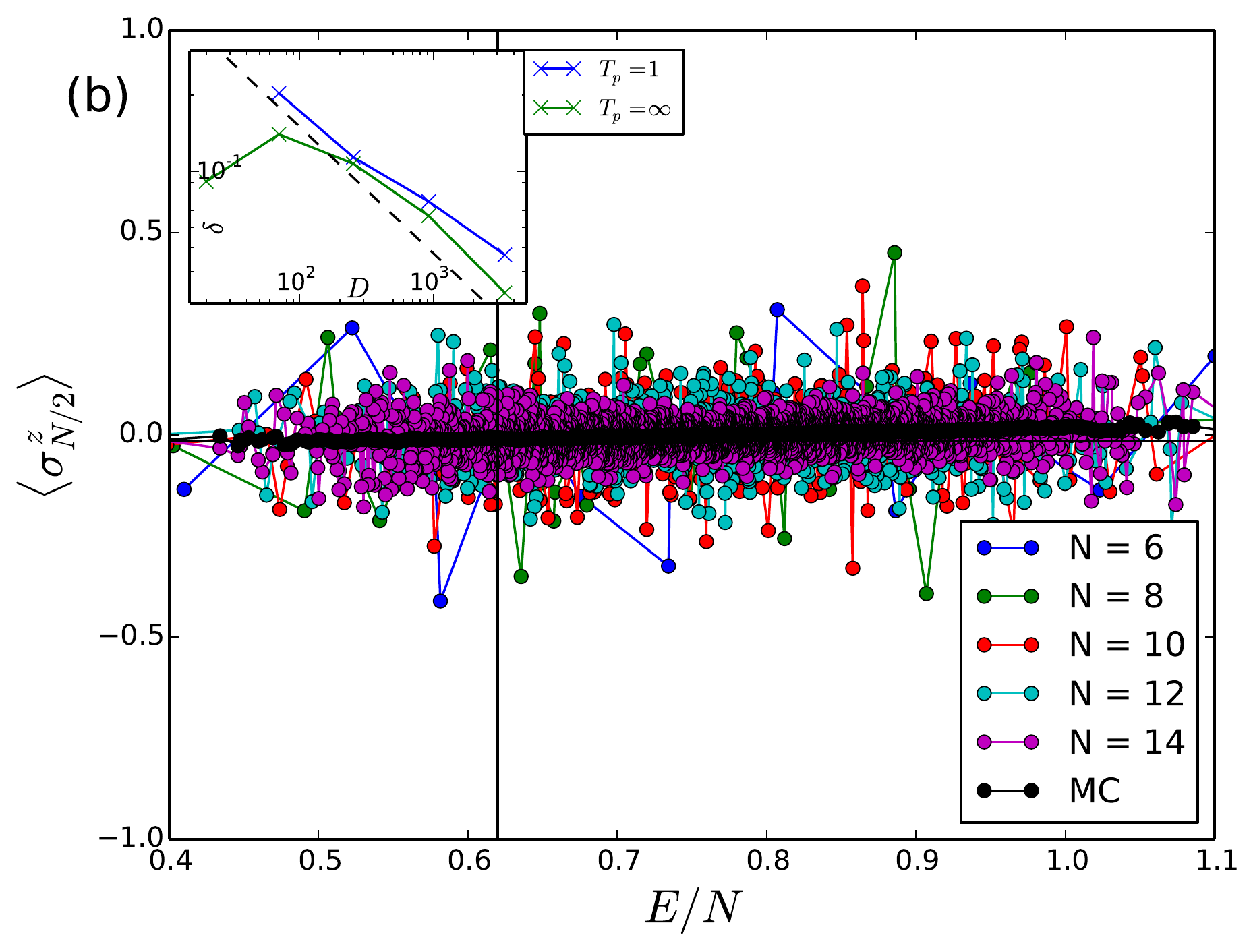}\\
\includegraphics[scale = 0.45]{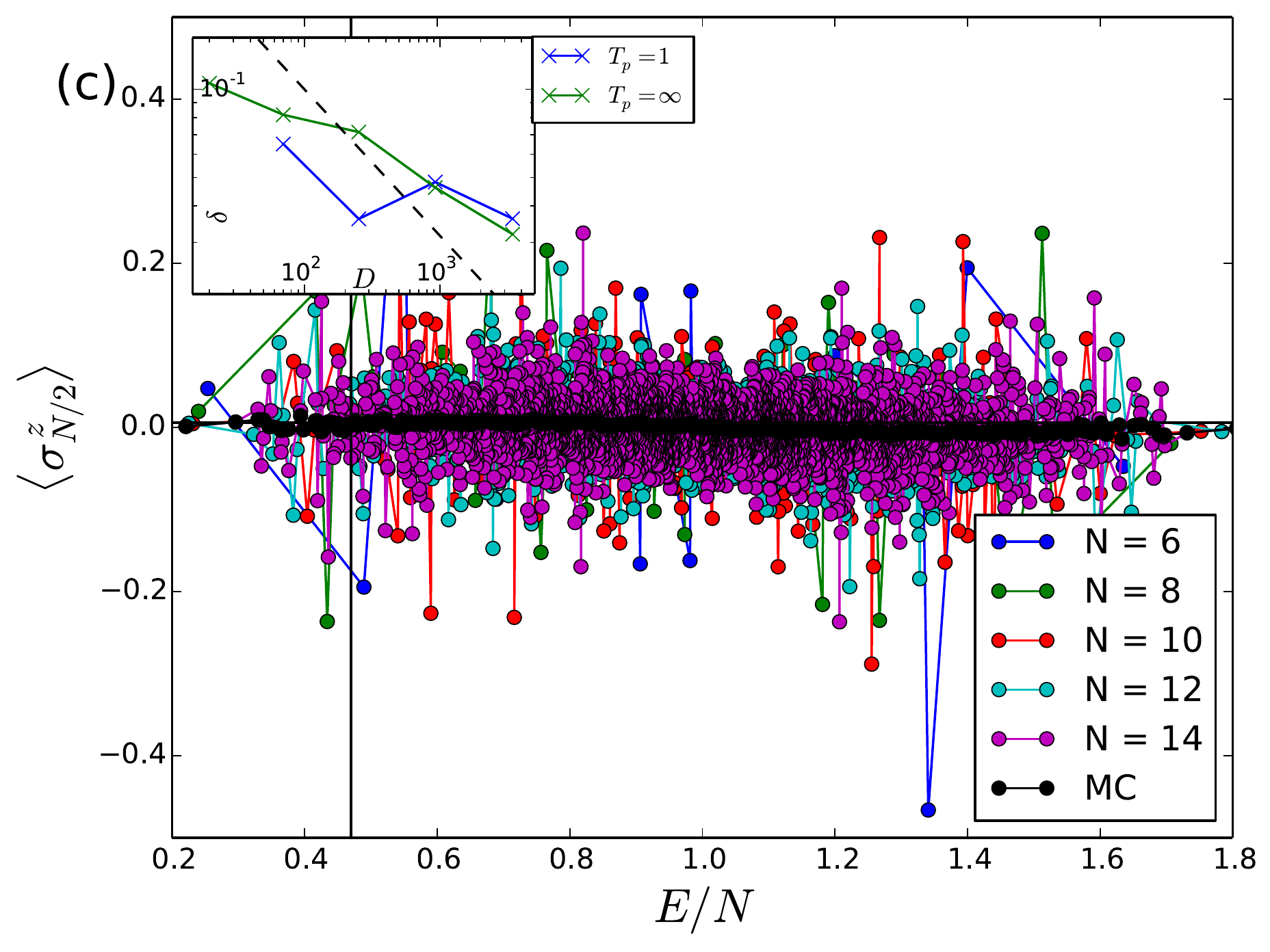}&\includegraphics[scale = 0.45]{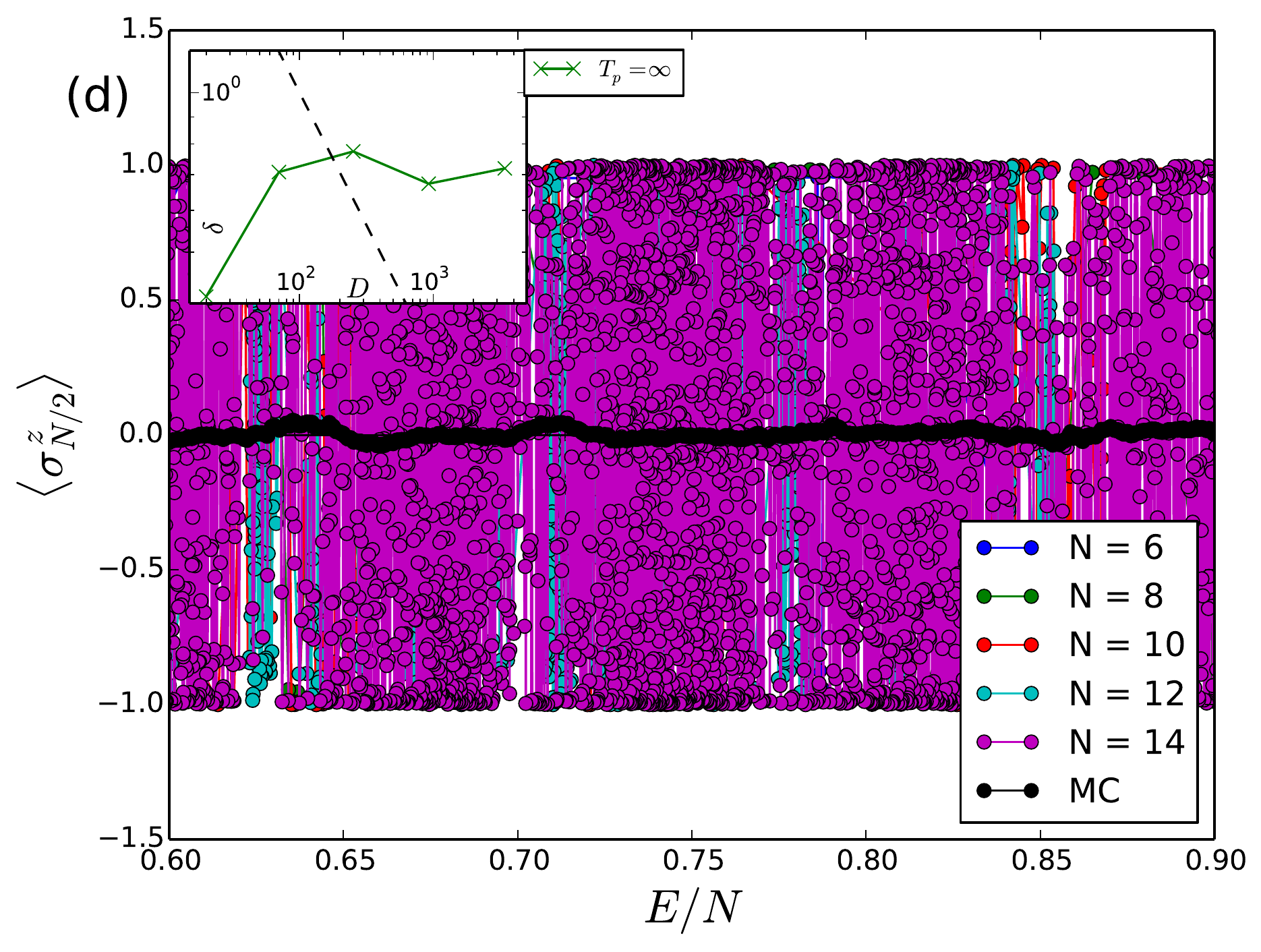}
\end{tabular}
\caption{(Color online) The expectation value and standard deviation of an operator $\sigma^z_{N/2}$ in the eigenstates of $\rhoss$ when the bulk Hamiltonian is (a) an XXZ model with a staggered magnetic field ($h_j = (-1)^j\times 0.5$, non-integrable), (b) an XXZ model without a magnetic field ($h_j = 0$, Bethe-Ansatz integrable), (c) a maximally driven XXZ model without a magnetic field ($(\mu, \bar{\mu}, \Delta, h_j) = (1.0, 0.0, 0.54, 0.0)$, exactly solvable $\rhoss$), (d) an XXZ model with a random magnetic field ($h_j \in [-10, 10]$, many-body localized). Unless otherwise mentioned, the data is shown for parameters $(\Delta, \Gamma, \mu, \bar{\mu}) = (0.5, 1, 0.5, 0.025)$ in the representative sector $S_z = 0$. See caption of Fig.~\ref{fig:inteth} for details on the labels.}
\label{fig:XXZeth}
\end{figure*}
In the previous section we showed evidence for the existence of NESS-ETH in the eigenstates of $\rhoss$ in the driven tilted Ising model as an example of a system with a non-integrable/chaotic Hamiltonian and just one conserved quantity. 
We now move on to the exploration of NESS-ETH in $\rhoss$ under less chaotic circumstances, such as the existence of additional conservation laws/symmetries, integrability, and localization.
A convenient and well-studied model to explore this physics is the driven XXZ model with a magnetic field, given by the Hamiltonian\cite{prosen2009matrix, prosen2010long, prosen2011open,  prosen2013eigenvalue, mendoza2013dephasing, vznidarivc2017dephasing, mendoza2018asymmetry}
\begin{equation}
    H_{\textrm{XXZ}} = \sumal{j = 1}{N-1}{\left(\sigma^x_j \sigma^x_{j+1} + \sigma^y_j \sigma^y_{j+1} + \Delta \sigma^z_j \sigma^z_{j+1}\right)} + \sumal{j = 1}{N}{h_j \sigma^z_j},
\label{eq:XXZmodel}
\end{equation}
with the jump operators
\begin{eqnarray}
    &&L_1 = \sqrt{\Gamma\left(1 - \mu + \bar{\mu}\right)}\sigma^+_1 \;\;\; L_2 = \sqrt{\Gamma\left(1 + \mu - \bar{\mu}\right)}\sigma^{-}_1 \nn \\
    &&L_3 = \sqrt{\Gamma\left(1 + \mu + \bar{\mu}\right)}\sigma^+_N \;\;\; L_4 = \sqrt{\Gamma\left(1 - \mu - \bar{\mu}\right)}\sigma^-_N.\nn \\
\label{eq:XXZLindblad}
\end{eqnarray}
$\mu$ indicates the driving strength of the system ($\mu = 0$ corresponds to equilibrium) and $\bar{\mu}$ is the average chemical potential.
To obtain $\rhoss$ for this system, we use the same methods as described in Sec.~\ref{subsec:testingethmodels}.
\subsection{Symmetries of $\rhoss$}
We now address symmetries that appear in $\rhoss$, assuming the existence of a unique steady state. 
When the steady state is unique, any symmetry of the Lindbladian is a symmetry of $\rhoss$.\cite{albert2014symmetries}
That is, as shown in App.~\ref{app:symmetries},
\begin{eqnarray}
    &U^\dagger H U = H,\;\;\; U^\dagger L_\mu U = \sumal{\nu}{}{\eta_{\mu\nu} L_\nu},\;\;\;\sumal{\mu}{}{\eta_{\mu\nu}^\ast \eta_{\mu\sigma}} = \delta_{\nu\sigma} \nn \\
    &\implies U^\dagger \rhoss U = \rhoss.
\label{eq:symmetryop}
\end{eqnarray}
For example, the Hamiltonian of Eq.~(\ref{eq:XXZmodel}) has a $U(1)$ symmetry, where
\begin{equation}
    U = \exp\left(i\phi S_z\right),\;\;\; S_z = \sumal{j = 1}{N}{\sigma^z_j}.
\label{eq:U1symmetry}
\end{equation}
Using $U$ of Eq.~(\ref{eq:U1symmetry}), it is straightforward to verify that Eq.~(\ref{eq:symmetryop}) is satisfied for the Hamiltonian and jump operators of Eqs.~(\ref{eq:XXZmodel}) and (\ref{eq:XXZLindblad}), and thus $S_z$ is a conserved quantity of $\rhoss$.
To obtain representative pure states in the presence of symmetries, representative values of all other conserved quantities need to be determined in addition to the representative pseudoenergy $\epsilon^\ast$.
For example, in the presence of $U(1)$ symmetry, in addition to Eq.~(\ref{eq:reppseudodensity}), the total spin density of the representative state is specified as
\begin{equation}
    s^\ast \equiv \frac{S_z^\ast}{N} =  \frac{\textrm{Tr}\left(S_z e^{-\hamilss} \right)}{N},
\label{eq:repspindensity}
\end{equation}
where $S_z^\ast$ is the $S_z$ quantum number of the representative state.
Note that the eigenstate of $\rhoss$ at pseudoenergy density $\epsilon^\ast$ and total spin density $s^\ast$ is a true representative state only if expectation values of local operators in states are smooth functions of pseudoenergy density $\epsilon$ and total spin density $s$.
Since testing their smoothness with $s$ is hard for the small systems we work with, we ensure that $s^\ast \approx 0$ in all the cases we explore so that representative states are always in the $S_z = 0$ sector for these system sizes. 
To test NESS-ETH in the presence of symmetries, we probe the Hamiltonian Eq.~(\ref{eq:XXZmodel}) with the jump operators of Eq.~(\ref{eq:XXZLindblad}) in a non-integrable regime, i.e. in the presence of a staggered magnetic field. 
Indeed, similar to the case for the tilted Ising model in Sec.~\ref{sec:testingeth}, we find that NESS-ETH holds for eigenstates within each quantum number sector of $S^z$.
The results for an operator in the bulk of the system in the sector $S_z = 0$ is shown in Fig.~\ref{fig:XXZeth}a.
Note that $\rhoss$ could in principle have symmetries that are not symmetries of the Lindbladian.\cite{albert2014symmetries}
While few-qubit systems with such symmetries can be explicitly constructed, we are not aware of any such examples in the context of boundary-driven systems.
\subsection{Integrable systems}
We now consider open quantum systems where the bulk Hamiltonian is integrable, and show that while the full NESS-ETH is violated, useful amounts of the structure still survive.  
We begin by noting that unlike for non-integrable systems, $\rhoss$ is now generally not close to the Gibbs state even when the left and right contacts are the same due to lack of ergodicity in the bulk of the system,\cite{prosen2008quantum, vznidarivc2010thermalization}.\footnote{Although the system can thermalize when the size of the contacts and the bulk are comparable.\cite{zanoci2016entanglement} }
Furthermore, the structure of $\rhoss$ depends on the exact structure jump operators on the edge, 
not just its physical parameters.\cite{vznidarivc2010thermalization}
Hence the status of NESS-ETH is not immediately apparent, even in equilibrium.
We discuss three cases in the following.
We start with the case where the bulk Hamiltonian is non-interacting, for example the XX model 
with a magnetic field (Eq.~(\ref{eq:XXZmodel}) with $\Delta = 0$).
The XX model can be written as a quadratic Hamiltonian in fermion operators after a Jordan-Wigner transform.\cite{lieb1961two}
The first subcase involves jump operators being linear in the fermion operators (which the operators of Eq.~(\ref{eq:XXZLindblad}) are after a Jordan-Wigner transform). Then the Lindbladian can be written as a ``quadratic superoperator".
Ref.~[\onlinecite{prosen2008third}] obtained a general method to obtain $\rhoss$ of such systems, which we have summarized in App.~\ref{app:prosen}.
As shown in Eqs.~(\ref{eq:nessexp}) and (\ref{eq:hamilssquad}), $\hamilss$ defined in Eq.~(\ref{eq:Hness}) can be written as a quadratic Hamiltonian.\footnote{The same result has been derived using alternate methods, for example in Ref.~[\onlinecite{banchi2014quantum}]}  
Thus, the eigenstates of $\rhoss$ do not generically satisfy NESS-ETH, as expected for non-interacting Hamiltonians and the spectrum exhibits Poisson statistics.
However, as shown in App.~\ref{app:restrictedeth}, certain eigenstates do obey a version of ``restricted NESS-ETH", \cite{khemani2014eigenstate} which allows the construction of representative pure states of $\rhoss$ for certain operators.
The second subcase has jump operators which are not linear in the fermion operators (for example $\sigma^z_1$). Here we find that $\hamilss$ still exhibits Poisson statistics. 
The standard deviations of typical operators show a $1/\sqrt{N}$ decay typical of non-interacting fermion Hamiltonians, presumably because $\hamilss$ in this case is quadratic with small quartic perturbations.  
The second case involves setting $\Delta \neq 0$ and in the presence of a uniform magnetic field ($h_j = h$), whereupon the XXZ model of Eq.~(\ref{eq:XXZmodel}) is integrable via the Bethe Ansatz.\cite{korepin1997quantum} 
Consequently, the energy levels of the Hamiltonian shows Poisson level statistics, typical of integrable models.\cite{poilblanc1993poisson}
However, the spectrum of $\hamilss$ shows GUE level statistics for a generic boundary driving of the XXZ model of Eq.~(\ref{eq:XXZmodel}) with a uniform magnetic field, even though the bulk Hamiltonian is integrable.\cite{prosen2013eigenvalue}
In Fig.~\ref{fig:XXZeth}b we show that operators in the bulk of the system satisfy NESS-ETH and the $1/\sqrt{D}$ scaling of the standard deviation of expectation values of operators, and representative pure states can be constructed.
The final case involves systems where $\rhoss$ can be analytically obtained in terms of a matrix product ansatz.\cite{prosen2011exact, prosen2014exact, karevski2013exact, prosen2015matrix}
An example is the maximally driven XXZ model of Eq.~(\ref{eq:XXZmodel}) with $\mu = 1$, $\bar{\mu} = 0$ in Eq.~(\ref{eq:XXZLindblad}).  
For a system size of $N$, $\rhoss$ can be written in terms of a Matrix Product Operator (MPO) of bond dimension $(N/2 + 1)^2$ (see Eq.~(13) of Ref.~[\onlinecite{prosen2011exact}]).
We find that the eigenvalues of $\hamilss$ show Poisson level statistics,\cite{prosen2013eigenvalue} and its eigenstates do not satisfy NESS-ETH.
In Fig.~\ref{fig:XXZeth}c, we show the expectation value of a local operator and its standard deviation for this case.
The existence of representative pure states is not clear for the system size $N$ we are able to test, although certain operators exhibit a power-law in $N$ decay of standard deviation, analogous to integrable Hamiltonians.\cite{khemani2014eigenstate, alba2015eigenstate}
Note that the connection between the solvability of $\rhoss$ and the solvability of its eigenstates is not fully clear, although a variation of the Algebraic Bethe Ansatz has been proposed to obtain eigenstates of $\rhoss$.\cite{prosen2013exterior, prosen2015matrix}
\subsection{Localization}
We briefly examine the case when the bulk Hamiltonian is Many-Body Localized (MBL). \cite{pal2010many, nandkishore2015many}
The XXZ Hamiltonian of Eq.~(\ref{eq:XXZmodel}) in the presence of a disordered field $h_j$ chosen uniformly from an interval $[-W, W]$ is known to be MBL for $W \geq W_c \approx 3.5-4.5$.\cite{pal2010many, luitz2015many, devakul2015early, doggen2018many}  
We observe that $\rhoss$ shows Poisson level statistics for $W = 10$ for $N = 12$, indicating that $\hamilss$ could be MBL.
Fig.~\ref{fig:XXZeth}d shows the expectation value of $\sigma^z_{N/2}$ in the bulk of the system for $W = 10$, where we clearly see the violation of NESS-ETH, which also rules out constructing representative states. 
However, we observe that the disorder strength $W_c$ required for the Poisson level statistics of $\rhoss$ strongly drifts with system size, and we are not able to exclude the possibility that the apparent MBL of $\hamilss$ is a finite-size effect. We plan to a more careful examination of
the behavior of bulk MBL systems under weak driving in future work. 
\section{Concluding Remarks}\label{sec:conclusions}
We have conjectured and given evidence for a generalization of ETH for isolated quantum systems to non-equilibrium steady states in open quantum systems, which we call the Non-Equilibrium Steady State Eigenstate Thermalization Hypothesis (NESS-ETH). 
We focused on boundary-driven one-dimensional systems described by GKLS master equations with local chaotic Hamiltonians. For such systems we find that the NESS density matrices $\rhoss$ qualitatively resemble equilibrium Gibbs density matrices in that (i) their level statistics show random matrix behavior for nearby levels and (ii) their eigenstates show smooth variation of the expectation values of few body operators with pseudoenergy density---the energy of $\hamilss$ defined in Eq.~(\ref{eq:Hness}), per site.
We showed that this smooth dependence makes it possible to pick representative pure states that reproduce the expectation values of few-body operators computed with the starting density matrices $\rhoss$.
We further showed that the density matrices of such representative states can be interpreted as weak solutions of the GKLS Master Equation. 

Some of these results continue to hold even when the Hamiltonians are integrable. For example, as we show in App.~\ref{app:restrictedeth}, free fermion GKLS systems introduced by Prosen exhibit a form of NESS-ETH for operators in real space similar to those for Hamiltonian systems albeit with a $1/\sqrt{N}$ decrease in the fluctuations and Poisson statistics.  For systems with additional symmetries we find that NESS-ETH is valid within each quantum number sector. When the bulk Hamiltonian is integrable, the level statistics is GUE or Poisson depending on whether the choice of jump operators allow an exact determination of $\rhoss$. NESS-ETH is violated in the latter case, although the construction of representative states might be possible for certain operators. For an MBL system we show that NESS-ETH breaks down entirely.

In future work it would be useful to test NESS-ETH for a wider class of systems. It will also be interesting to examine the behavior of off-diagonal matrix elements which we have not touched in this work. Finally, it is tempting to ask if the weak transport solutions presented herein can be related to strong solutions in some other formalism---analogs of scattering states in scattering theory.
%

%

%
%
%
%
%

%
%
%
%
\section*{Acknowledgements}
We thank Toma\v{z} Prosen for an introduction to his work and for useful discussions, Vedika Khemani for useful discussions and pointing out an error in an earlier version of the draft, and David Huse for comments on a draft.
TD is supported in part by the National Science Foundation under Grant No. NSF PHY-1748958. SLS is supported by US Department of Energy grant No. DE-SC0016244. This research was also supported by funding from the Defense Advanced Research Projects Agency (DARPA) under the DRINQS program. The views, opinions and/or findings expressed are those of the authors and should not be interpreted as representing the official views or policies of the Department of Defense or the U.S. Government. 

\appendix
\section{On the Validity of Eq.~(\ref{eq:rhonessapprox})}\label{app:validity}
The Schmidt decomposition of $\rho_G$ between the bulk $B$ and the contacts $C$ reads (suppressing the $\beta$-dependence) 
\begin{equation}
    \rho_G(\beta) = \sumal{\alpha}{}{\lambda_\alpha \rho^{(B)}_\alpha \otimes \rho^{(C)}_\alpha}, 
\label{eq:rhoGschmidt}
\end{equation}
where $\rho^{(B)}_\alpha$ and $\rho^{(C)}_\alpha$ are supported on the bulk and the contacts respectively. 
If $\rhoss = \rho_G(\beta)$, it must satisfy the GKLS Master Equation of Eq.~(\ref{eq:lindbladeq}).  
Since $\hmL_B\left(\rho_G\right) = 0$, we obtain
\begin{eqnarray}
    &&\hmL \left(\rho_G\right) = \hmL_B \left(\rho_G\right) + \hmL_C \left(\rho_G\right) = 0 \nn \\
    &&\implies  \sumal{\alpha}{}{\lambda_\alpha \hmL_C \left(\rho^{(B)}_\alpha \otimes \rho^{(C)}_\alpha\right)} = 0 \nn \\
    &&\implies \sumal{\alpha}{}{\lambda_\alpha \rho^{(B)}_\alpha \otimes \hmL_C \left(\rho^{(C)}_\alpha\right)} = 0 \nn \\
    &&\implies \hmL_C\left(\rho^{(C)}_\alpha\right) = 0 \;\;\; \forall \alpha,
\label{eq:contradiction}
\end{eqnarray}
where we have used that the set $\{\rho^{(B)}_\alpha\}$, by virtue of being the Schmidt vectors, is linearly independent.  
Since $\rho_G$ generally has a full-rank over any Schmidt decomposition, Eq.~(\ref{eq:contradiction}) would imply $\hmL_C\left(\cdot\right) \equiv 0$, which is a contradiction. 
However, $\rho_G$ is known to admit an efficient Matrix Product Operator (MPO) representation. 
Equivalently, the tails of the distribution of the Schmidt values $\{\lambda_\alpha\}$ in Eq.~(\ref{eq:rhoGschmidt}) fall off exponentially.
It is reasonable to expect that Eq.~(\ref{eq:rhonessapprox}) holds approximately, where Eq.~(\ref{eq:contradiction}) need not hold for $\alpha$'s for which $\lambda_\alpha \approx 0$. 
\section{Structure of jump operators}\label{app:jump}
We start with Eq.~(\ref{eq:reqdbaths}), and with $\rho_G\left(\beta\right)$ defined as
\begin{equation}
    \rho_G\left(\beta\right) = Z^{-1}\exp\left(-\beta\left(h\sigma^x + g\sigma^z\right)\right),
\label{eq:rhoGorig}
\end{equation}
where $\left\{\sigma^\alpha\right\}$ are the Pauli matrices.
For ease of calculation, we write $\rho_G\left(\beta\right)$ as (dropping the $\beta$ dependence)
\begin{eqnarray}
    &\rho_G = Z^{-1} \exp\left(-\beta \sqrt{g^2 + h^2}\left(\frac{h}{\sqrt{g^2 + h^2}}\sigma^z + \frac{g}{\sqrt{g^2 + h^2}}\sigma^z\right)\right) \nn \\
    &= Z^{-1} \exp\left(-\beta \sqrt{g^2 + h^2}\left(\sigma^z\cos\theta + \sigma^x\sin\theta\right)\right) \nn \\
    &\equiv Z^{-1}\exp\left(-\beta m \tau^z\right),
\label{eq:rhoG}
\end{eqnarray}
and we've defined
\begin{eqnarray}
    &\cos\theta \equiv \frac{h}{\sqrt{g^2 + h^2}},\;\;\sin\theta \equiv \frac{g}{\sqrt{g^2 + h^2}},\;\;m \equiv \sqrt{g^2 + h^2},\nn \\
    &\tau^z \equiv \sigma^z\cos\theta + \sigma^x\sin\theta. 
\label{eq:defns}
\end{eqnarray}
We choose the one-site jump operators
\begin{equation}
    L_1 = \sqrt{\Gamma_1}\tau^+,\;\; L_2 = \sqrt{\Gamma_2}\tau^-. 
\end{equation}
With this choice of jump operators, we want to solve Eq.~(\ref{eq:reqdbaths}), which reads
\begin{eqnarray}
    &\sumal{\mu = 1}{2}{\left(2 L_\mu \rho_G L^\dagger_\mu - L^\dagger_\mu L_\mu \rho_G - \rho_G L^\dagger_\mu L_\mu\right)} = 0 \nn \\ 
    &\sumal{\alpha \in \{+, -\}}{}{\Gamma_\alpha \left(2 \tau^\alpha e^{-\beta m \tau^z} \left(\tau^\alpha\right)^\dagger - \left(\tau^\alpha\right)^\dagger \tau^\alpha e^{-\beta m \tau^z}\right.} \nn \\
    &{\left.- e^{-\beta m \tau^z} \left(\tau^\alpha\right)^\dagger \tau^\alpha\right)} = 0 \nn \\
\label{eq:taucondition}
\end{eqnarray}
It is straightforward to check that Eq.~(\ref{eq:taucondition}) is true as long as
\begin{equation}
    \beta = -\frac{1}{2m}\log\left(\frac{\Gamma_1}{\Gamma_2}\right), 
\label{eq:betaexpression}
\end{equation}
which is Eq.~(\ref{eq:tiltedtemp}).
To express $\tau^\alpha$'s in terms of $\sigma^\alpha$'s, note that
\begin{equation}
    \tau^z = R^\dagger \sigma^z R,
\end{equation}
where we define $R$ as
\begin{equation}
    R = \exp\left(i\frac{\theta}{2}\sigma^y\right) = 
    \begin{pmatrix}
        \cos\left(\frac{\theta}{2}\right) & \sin\left(\frac{\theta}{2}\right) \\
        -\sin\left(\frac{\theta}{2}\right) & \cos\left(\frac{\theta}{2}\right)
    \end{pmatrix}.
\label{eq:Rdefn}
\end{equation}
Thus we obtain
\begin{equation}
    \tau^{\pm} = R^\dagger \sigma^{\pm} R = \frac{1}{2}\left(-\sigma^z\sin\theta + (1 + \cos\theta)\sigma^{\pm} - (1-\cos\theta)\sigma^{\mp}\right).
\label{eq:taupm}
\end{equation}
where we use the following definitions of the $\sigma^{\pm}$ matrices
\begin{eqnarray}
    &\sigma^+ = \frac{1}{2}\left(\sigma^x + i\sigma^y\right) = 
    \begin{pmatrix}
    0 & 1 \\
    0 & 0
    \end{pmatrix},\nn \\
    &\sigma^- = \frac{1}{2}\left(\sigma^x - i\sigma^y\right) = 
    \begin{pmatrix}
        0 & 0 \\
        1 & 0
    \end{pmatrix}.
\label{eq:sigmapmdefn}
\end{eqnarray}
\section{TEBD with one-site jump operators}\label{app:tebddetails}
In this appendix, we describe a method for obtaining the steady state density matrix using 
The steady state density matrix $\rhoss$ is obtained by simply time evolving an initial density matrix using the GKLS master equation until the steady state is reached.
We represent the density matrix as a pure state on a doubled Hilbert space $\mathcal{H} \otimes \mathcal{H}$ composed of two copies of the original Hilbert space and use the standard TEBD algorithm to implement time evolution.\cite{vidal2004efficient}
A general density matrix $\rho$ may always be purified by introducing additional degrees of freedom
\begin{equation}
    \ket{\rho} \equiv \sum_{n m} \bra{n}{\rho}\ket{m}\ket{n}_{A} \otimes \ket{m}_{B}
\end{equation}
where $\{\ket{n}_A\}$ and $\{\ket{n}_B\}$ are a complete basis for each of the two copies of the Hilbert space $\mathcal{H}$, which we denote by $\mathcal{H}_A$ and $\mathcal{H}_B$ respectively, which together form a complete basis for the doubled Hilbert space $\mathcal{H}_A \otimes \mathcal{H}_B = \mathcal{H} \otimes \mathcal{H}$.
In this language, the Lindbladian acts as an operator as
\begin{align}
\begin{split}
\hmL  =& -i(H_A - H_B^{T}) \\
        &+ \sum_\mu \left( 2 L_{\mu, A} L_{\mu, B}^{*} - L^\dagger_{\mu, A} L_{\mu, A} - L^{T}_{\mu, B} L_{\mu, B}^{*} \right)
\end{split}
\end{align}
where subscripts $A$ and $B$ indicate the operators acting on the Hilbert space $\mathcal{H}_{A}$ and $\mathcal{H}_B$.
The GKLS master equation of Eq.~(\ref{eq:lindbladeq}) is given by
\begin{equation}
    \frac{d}{dt}\ket{\rho} = \hmL \ket{\rho}
\end{equation}
which is solved for an initial condition $\ket{\rho(0)}$ by
\begin{equation}
\ket{\rho (t)} = e^{\hmL t}\ket{\rho(0)}
\end{equation}
For a Hamiltonian $H$ with two-site interaction terms, coupling to a bath on the left and right edges via single-site jump operators $L_\mu$ at either edge, we may decompose the master equation Eq.~(\ref{eq:lindbladeq}) as
\begin{align}
    \hmL = \hmL_\mathrm{even} + \hmL_\mathrm{odd}
\end{align}
where $\hmL_\mathrm{even (odd)}$ are each composed of a sum of terms with disjoint support.
In our case, $\hmL_\mathrm{even}$ contains only terms from $H$ which couple between sites $2j$ and $2j+1$ on the $N$-site chain, and if $N$ is odd, it includes the terms from $\hmL_{C_r}$ (see Eq.~(\ref{eq:LCleftright})) which act on site $N$.
$\hmL_\mathrm{odd}$ contains terms from $H$ which couple between sites $2j+1$ and $2j+2$, the terms in $\hmL_{C_l}$ which have support on site $1$, and provided $N$ is even, the terms from $\hmL_{C_r}$ terms which act on site $N$.
In this way, we write $\hmL_\mathrm{even}$ and $\hmL_\mathrm{odd}$ that have distinct supports.
We then proceed by using the Suzuki-Trotter decomposition\cite{vidal2004efficient} for the time evolution by a small time step $\delta t$,
\begin{equation}
    e^{\hmL \delta t} \approx e^{\hmL_\mathrm{even} \delta t}
     e^{\hmL_\mathrm{odd} \delta t} + \mathcal{O}(\delta t^2).
\end{equation}
The density matrix $\ket{\rho}$ is represented as a matrix product state (MPS) with two physical indices per site (one each for the Hilbert spaces $\mathcal{H}_A$ and $\mathcal{H}_B$) whereas the Lindbladian $\hmL$ as a matrix product operators (MPO) with two sets of physical indices per site.  
The simulation begins with an initial state $\ket{\rho(0)}$, which we choose to be the purified infinite-temperature density matrix with an MPS bond dimension 1.
Using the standard TEBD algorithm, $e^{\hmL \delta t}$ is applied to the MPS, while keeping the bond dimension below 50.
We begin with an initial time-step $\delta t = 1$.
The entanglement entropy of $\ket{\rho(t)}$ (with respect to a bipartition in the middle of the system) is used to determine the convergence of the state, and the time-step is updated as $\delta t \rightarrow 0.9 \delta t$ as $\ket{\rho(t)}$ converges. 
The evolution of the state is evolved $\ket{\rho(t)}$ reaches a steady state $\ket{\rhoss}$, at which point the full $2^N$-dimensional matrix representation $\rhoss$ can easily be obtained by contracting the MPS for $\ket{\rhoss}$.
We note that since the overall norm of the wavefunction is preserved in the TEBD algorithm, the resulting density matrix satisfies $\braket{\rhoss}{\rhoss} = 1$ rather than $\textrm{Tr}\left(\rhoss\right) = 1$, and thus needs to be scaled by an overall factor to obtain the normalized density matrix. 
\section{General symmetries of \boldsymbol{$\rhoss$}}\label{app:symmetries}
In this appendix, we provide a proof of Eq.~(\ref{eq:symmetryop}).
The NESS is the fixed point of the GKLS master equation of Eq.~(\ref{eq:lindbladeq}) ($\hmL\left(\rhoss\right) = 0$), and thus
\begin{equation}
    -i\comm{H}{\rhoss} + \sum_{\mu}{\left(2 L_\mu \rhoss L^\dagger_\mu - \anticomm{L^\dagger_\mu L_\mu}{\rhoss}\right)} = 0.
\label{eq:LMEsym1}
\end{equation}
Using the conditions of Eq.~(\ref{eq:symmetryop}), Eq.~(\ref{eq:LMEsym1}) can be written as
\begin{eqnarray}
    &-i\comm{U H U^\dagger}{\rhoss} + \sumal{\mu, \nu, \sigma}{}{\eta_{\mu\nu}^\ast \eta_{\mu\sigma}\left(2 U L_\sigma U^\dagger \rhoss U L^\dagger_\nu U^\dagger\right.} \nn \\
    &\left.- \anticomm{U L^\dagger_\nu L_\sigma U^\dagger}{\rhoss}\right) = 0.
\label{eq:LMEsym2}
\end{eqnarray}
Left- and right-multiplying by $U^\dagger$ and $U$ respectively and using the property of $\eta_{\mu\nu}$ in Eq.~(\ref{eq:symmetryop}), we obtain
\begin{equation}
    -i\comm{H}{U^\dagger \rhoss U} + \sum_{\mu}{\left(2 L_\mu U^\dagger \rhoss U L^\dagger_\mu - \anticomm{L^\dagger_\mu L_\mu}{U^\dagger \rhoss U}\right)} = 0.
\label{eq:LMEsymfin}
\end{equation}
Provided the NESS is unique, Eq.~(\ref{eq:LMEsymfin}) implies
\begin{equation}
    U^\dagger \rhoss U = \rhoss.
\label{eq:rhosssymapp}
\end{equation}
\section{Structure of \boldsymbol{$\rhoss$} in quadratic systems}\label{app:prosen}
Here we review the structure of $\rhoss$ in quadratic fermion systems using the formalism developed in Ref.~[\onlinecite{prosen2008third}] and show that $\hamilss$ is also quadratic in such cases. Note that the same result has appeared the literature in other contexts,\cite{banchi2014quantum} we include this section here for emphasis and completeness.  
For a chain of $N$ sites, we consider the Hamiltonian and jump operators to be of the form
\begin{eqnarray}
    H &=& \sum_{j,k = 1}^{2N}{w_j H_{jk} w_k} \nn \\
    L_\mu &=& \sum_{j = 1}^{2N}{l_{\mu,j} w_j},
\label{eq:hamillindbladform}
\end{eqnarray}
where $\{w_j\}$ are the Majorana fermions that obey the algebra
\begin{equation}
    \anticomm{w_j}{w_k} = 2\delta_{j,k},\;\;\; j,k = 1, 2, \dots, 2N.
\label{eq:majoranaalg}
\end{equation}
To solve the GKLS equation of Eq.~(\ref{eq:lindbladeq}) with the operators of Eq.~(\ref{eq:hamillindbladform}), we define a Fock space of operators $\{\ket{P_{\vec{\alpha}}}\}$, where
\begin{equation}
    \vec{\alpha} = \left(\alpha_1, \alpha_2, \cdots, \alpha_{2N}\right).
\end{equation}
The operators $\{P_{\vec{\alpha}}\}$ in this space read
\begin{equation}
    P_{\left(\alpha_1, \alpha_2, \cdots, \alpha_{2N}\right)} \equiv w_1^{\alpha_1} w_2^{\alpha_2} \cdots w_{2N}^{\alpha_{2N}}, \;\;\; \alpha_j \in \{0,1\}.
\label{eq:opspace}
\end{equation}
In this operator space, we define adjoint creation and annihilation (super-)operators $\{\cdh_j\}$ and $\{\ch_j\}$ that ``create" and ``annihilate" Majorana operators $w_j$.
That is, their actions read
\begin{equation}
    \cdh_j \ket{P_{\vec{\alpha}}} = \delta_{\alpha_j, 0} \ket{w_j P_{\vec{\alpha}}},\;\;\; \ch_j \ket{P_{\vec{\alpha}}} = \delta_{\alpha_j, 1} \ket{w_j P_{\vec{\alpha}}}.
\label{eq:adjointfermionaction}
\end{equation}
Furthermore, using Eqs.~(\ref{eq:majoranaalg}) and (\ref{eq:adjointfermionaction}), we obtain that $\cdh_j$ and $\ch_j$ are fermion operators, i.e. they satisfy
\begin{equation}
    \{\ch_j, \cdh_k\} = \delta_{j,k},\;\{\ch_j, \ch_k\} = 0,\; j,k = 1, 2, \cdots, 2N. 
\label{eq:fermionalg}
\end{equation}
We further introduce Majorana superoperators that read
\begin{equation}
    \ah_{2j-1} = \frac{1}{\sqrt{2}}\left(\ch_j + \cdh_j\right),\;\;\; \ah_{2j} = \frac{i}{\sqrt{2}}\left(\ch_j - \cdh_j\right),\;\;\;
\end{equation}
such that
\begin{equation}
    \{\ah_r, \ah_s\} = \delta_{r,s}.
\end{equation}
In terms of these Majorana superoperators, the Lindbladian superoperator (for the even parity sector) can be written as\cite{prosen2008third} 
\begin{equation}
    \hmL = \sumal{j, k = 1}{4N}{A_{jk} \ah_j \ah_k} - A_0,
\label{eq:liouvillianquad}
\end{equation}
where\cite{prosen2008third}
\begin{eqnarray}
    &A_{2j-1, 2k-1} = -2i H_{jk} - M_{jk} + M_{kj},\;\; A_{2j-1, 2k} = 2i M_{kj},\nn \\
    &A_{2j, 2k -1} = -2i M_{jk},\;\; A_{2j, 2k} = -2i H_{jk} + M_{jk} - M_{kj}, \nn \\
\end{eqnarray}
where
\begin{equation}
    M_{jk} = \sumal{\mu}{}{l_{\mu, j} l^\ast_{\mu, k}}, \;\;\; A_0 = 2 \sumal{j = 1}{2N}{M_{jj}}.
\end{equation}
When diagonalized, the Lindbladian reads
\begin{equation}
    \hmL = -2\sumal{j = 1}{2N}{\beta_j \bph_j \bh_j},
\label{eq:liouvNMM}
\end{equation}
where $\{\bh_j\}$ and $\{\bph_j\}$ are the Normal Master Modes (NMMs) that read
\begin{equation}
    \bh_j = \sumal{k = 1}{4N}{v_{2j-1, k} \ah_k},\;\;\;\bh_j = \sumal{k = 1}{4N}{v_{2j, k} \ah_k},
\label{eq:NMMdefn}
\end{equation}
and obey fermion commutation relations
\begin{equation}
    \{\bh_j, \bph_k\} = \delta_{j,k},\;\{\bh_j, \bh_k\} = 0,\;j,k = 1,2,\cdots, 2N.
\label{eq:NMMcommutation}
\end{equation}
In Eq.~(\ref{eq:NMMdefn}), $\{v_{j,k}\}$ are the elements of the matrix $V$ that satisfies
\begin{equation}
    A = V^T \Lambda V, 
\end{equation}
where
\begin{equation}
    \Lambda = \bigoplus_{j = 1}^{2N}{
    \begin{pmatrix}
    0 & \beta_j \\
    -\beta_j & 0 
    \end{pmatrix}},
\end{equation}
where $\{\beta_j\}$ is the set of eigenvalues of $A$ arranged such that $\textrm{Re}\ \beta_1 \geq \textrm{Re}\ \beta_2 \geq \cdots \geq \textrm{Re}\ \beta_{2N} \geq 0$.
Note the eigenvalues appear in pairs $\left(\beta_j, -\beta_j\right)$ because $A$ is antisymmetric.
The NESS is then given by the ``Bogoliubov" ground state of the ``superconducting" non-Hermitian system given by Eq.~(\ref{eq:liouvillianquad}), and can be written as
\begin{equation}
    \ket{\rhoss} = \prodal{k = 1}{2N}\bh_k\ket{\mathds{1}}.
\label{eq:nessbogoliubov}
\end{equation}
Eqs.~(\ref{eq:liouvNMM}) and (\ref{eq:NMMcommutation}) then ensure that
\begin{equation}
    \hmL \ket{\rhoss} = 0. 
\end{equation}
To obtain the operator form of $\rhoss$, we first express $b_k$ in terms of $\ch_j$'s and $\cdh_j$'s as
\begin{equation}
    \bh_k = \sumal{j = 1}{2N}{\left(\Gamma_{kj} \ch_j + \Upsilon_{kj} \cdh_j\right)}.
\label{eq:NMMexpression}
\end{equation}
Since Eq.~(\ref{eq:nessbogoliubov}) has the form of a BCS wave function, the (unnormalized) NESS can be written as an exponential
\begin{equation}
    \ket{\rhoss} = \exp\left(-\frac{1}{2}\sumal{i,j = 1}{2N}{G_{ij} \cdh_i \cdh_j}\right)\ket{\mathds{1}},
\label{eq:nesspairing}
\end{equation}
where, following the appendix in Ref.~[\onlinecite{chung2001density}], the ``pairing function" $G_{ij}$ is given by the solution to the equation
\begin{equation}
    \sumal{k = 1}{2N}{\left(-\Gamma_{mk} G_{kn} + \Upsilon_{mn}\right)} = 0.
\label{eq:pairingfunction}
\end{equation}
Using Eqs.~(\ref{eq:adjointfermionaction}) and (\ref{eq:fermionalg}), the operator form of $\rhoss$ can be written as
\begin{equation}
    \rhoss = \exp\left(-i\sumal{i,j = 1}{2N}{K_{ij} w_i w_j}\right),
\label{eq:nessexp}
\end{equation}
and the NESS Hamiltonian $\hamilss$ defined in Eq.~(\ref{eq:Hness}) reads 
\begin{equation}
    \hamilss = i\sumal{i,j = 1}{2N}{K_{ij} w_i w_j}.
\label{eq:hamilssquad}
\end{equation}
To obtain the matrix $K$ in terms of $G$, we start with the computation of the off-diagonal elements $C_{ij}$ of the covariance matrix of $\rhoss$, defined as
\begin{equation}
    C_{ij} = \langle w_i w_j \rangle_{\texttt{ss}} = \textrm{Tr}\left(\rhoss w_i w_j\right).
\label{eq:covariancedefn}
\end{equation}
First, we compute $C$ assuming Eq.~(\ref{eq:nessexp}).
Since $i K$ is an antisymmetric matrix, it admits a spectral decomposition
\begin{equation}
    i K = R \Sigma R^T, 
\end{equation}
where $R^T R = \mathds{1}$ and $\Sigma$ is a block-diagonal matrix that reads
\begin{equation}
    \Sigma = \bigoplus_{l = 1}^{N}{
    \begin{pmatrix}
        0 & i\kappa_l \\
        - i\kappa_l & 0
    \end{pmatrix}},
\label{eq:sigmaform}
\end{equation}
$\{\kappa_l\}$ being the eigenvalues of $i K$. 
We then define Majorana fermions $\{\eta_l\}$ as
\begin{equation}
    \eta_l = \sumal{k = 1}{2N}{\left(R^T\right)_{lk} w_k}.
\label{eq:etadefn}
\end{equation}
In terms of the $\eta$ Majorana fermions, $\hamilss$ of Eq.~(\ref{eq:hamilssquad}) reads
\begin{equation}
    \hamilss = \sumal{k,l = 1}{2N}{\Sigma_{kl} \eta_k \eta_l} = 2i \sumal{l = 1}{2N}{\kappa_l \eta_{2l -1}\eta_{2l}},
\label{eq:hamilssdiag}
\end{equation}
where we have used Eqs.~(\ref{eq:sigmaform}) and (\ref{eq:etadefn}).
Further defining complex fermion operators $d_l$ and $d_l^\dagger$ out of the $\eta$ Majorana fermions, we can write $\hamilss$ and $\langle d^\dagger_l d_l \rangle_{\texttt{ss}}$ as
\begin{equation}
    \hamilss = \sumal{l = 1}{N}{ \kappa_l\left(2 d^\dagger_l d_l - 1\right)},\;\langle d^\dagger_l d_l \rangle_{\texttt{ss}} = \frac{1}{e^{\kappa_l} + 1}.
\end{equation}
Further, we obtain
\begin{equation}
    \langle\eta_{2l-1}\eta_{2l}\rangle_{\texttt{ss}} =  -2 i \langle d^\dagger_l d_l\rangle_{\texttt{ss}} + i = i\tanh\left(\frac{\kappa_l}{2}\right).
\end{equation}
The covariance matrix element of Eq.~(\ref{eq:covariancedefn}) can be written as 
\begin{eqnarray}
    C_{ij} &=& \langle w_i w_j \rangle_{\texttt{ss}} = \sumal{k, l = 1}{2N}{R_{ik}R_{jl}\langle \eta_k \eta_l \rangle_{\texttt{ss}}} \nn \\
    &=& \sumal{l = 1}{N}{\left(R_{i,2l-1} R_{j,2l} - R_{i,2l} R_{j,2l-1} \right)\langle \eta_{2l-1} \eta_{2l}\rangle_{\texttt{ss}}} \nn \\
    &=& \sumal{l = 1}{N}{\left(R_{i,2l-1} R_{j,2l} - R_{i,2l} R_{j,2l-1} \right)i \tanh\left(\frac{\kappa_l}{2}\right)} \nn \\
    &=& i \left(R \tanh\left(\frac{\Sigma}{2}\right) R^T\right)_{ij} = \left( \tanh\left(\frac{i K}{2}\right)\right)_{ij}.
\label{eq:covarianceusual}
\end{eqnarray}
Now we compute $C_{ij}$ using the superoperator formalism, which can be written as
\begin{eqnarray}
    C_{ij} &=& \langle w_i w_j \rangle_{\texttt{ss}} = \bra{\mathds{1}} \ch_i \ch_j \ket{\rhoss} \nn \\
    &=& \bra{\mathds{1}} \ch_i \ch_j \exp\left(-\frac{1}{2}\sumal{i,j = 1}{2N}{G_{ij} \cdh_i \cdh_j}\right)\ket{\mathds{1}} \nn \\
    &=& G_{ij}.
\label{eq:covariancesuperop}
\end{eqnarray}
Thus, using Eqs.~(\ref{eq:covarianceusual}) and (\ref{eq:covariancesuperop}), the $2N \times 2N$ matrices $i K$ and $G$ can be shown to be related according to
\begin{eqnarray}
    i K &=&  2\textrm{ arctanh}\left(G\right) \nn \\
    &=&  2\left(G + \frac{G^3}{3} + \frac{G^5}{5} + \frac{G^7}{7} + \dots\right).
\label{eq:nesshamil}
\end{eqnarray}
\begin{figure*}[ht!]
\centering
 \begin{tabular}{cc}
\includegraphics[scale = 0.45]{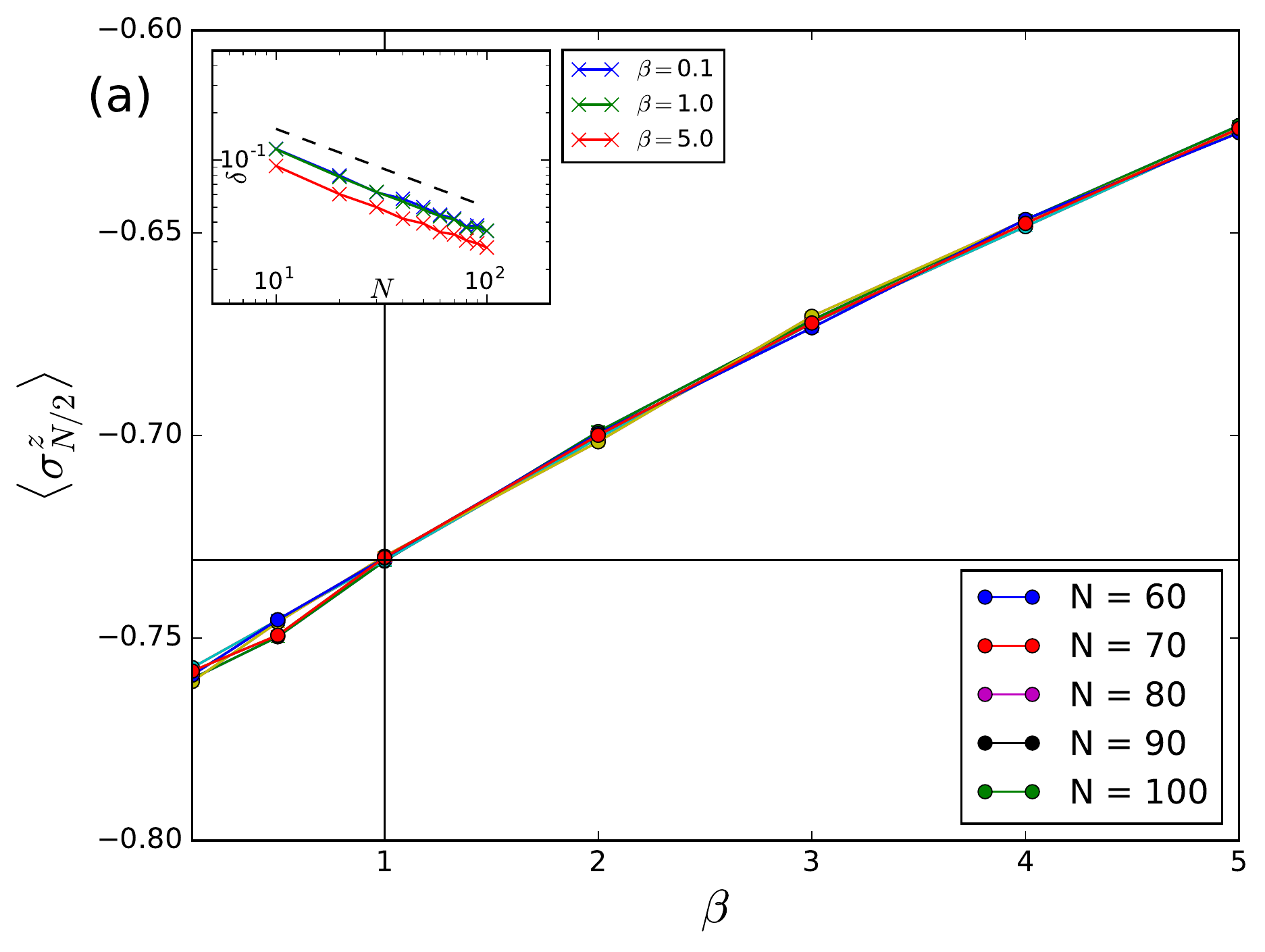}&\includegraphics[scale = 0.45]{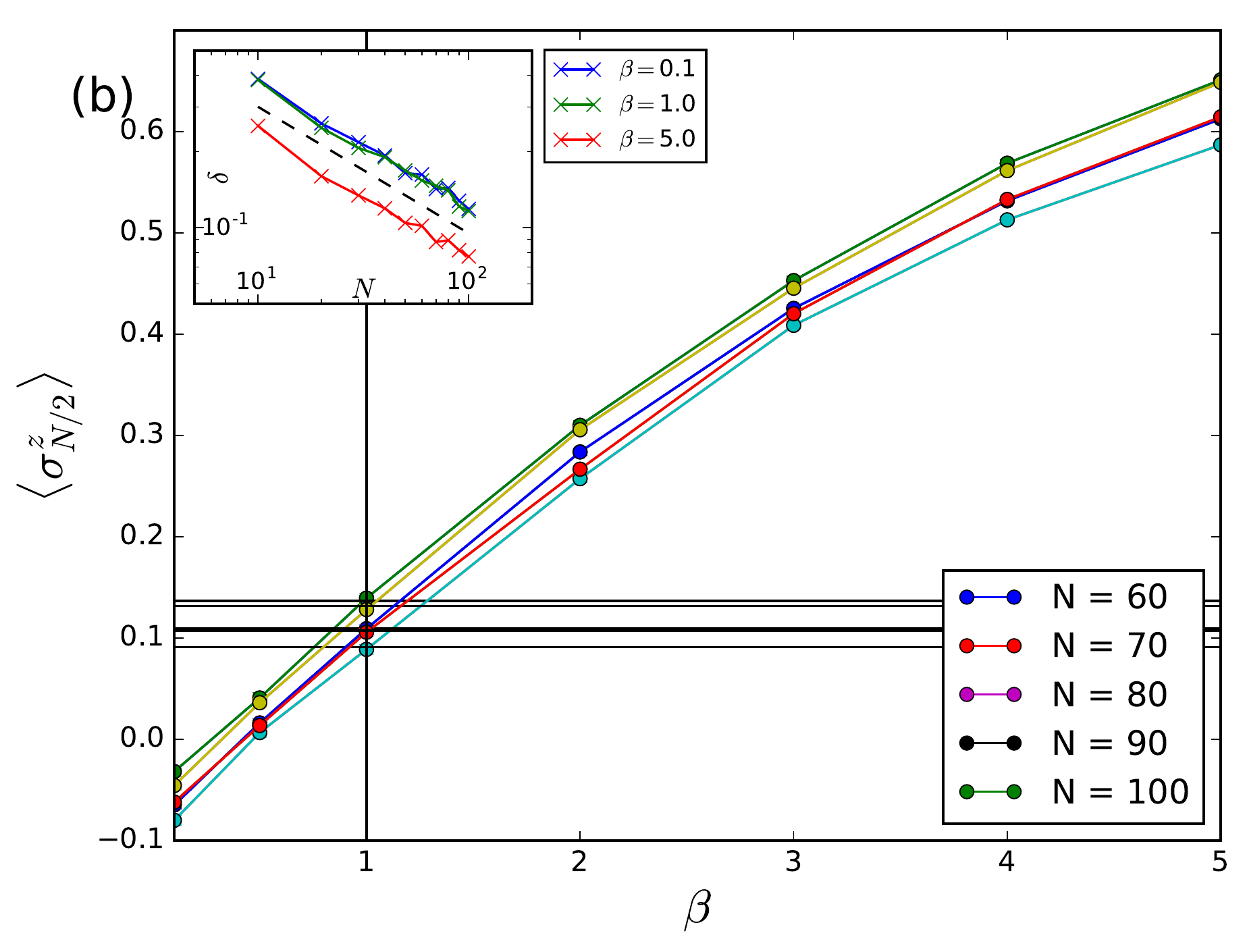}
\end{tabular}
\caption{(Color online) Expectation values and standard deviations of the operators $\sigma^z_{N/2}$ in the many-body eigenstates of $\rhoss$ of the XY model whose occupancies obey the Fermi-Dirac distribution with inverse pseudotemperature $\beta$. (a) $\rhoss$ has exhibits short-range correlations ($h = \gamma = 0.9$). (b) $\rhoss$ exhibits long-range correlations ($h = \gamma = 0.1$). (Main) The expectation value is a smooth function of $\beta$. The horizontal lines denote the expectation values of the operators in $\rhoss$ for various system sizes, which correspond to $\beta = 1$.  (Inset) Standard deviations decay as $~1/\sqrt{N}$ (denoted by the dotted line) for various $\beta$.  The jump operators used are of the form of Eq.~(\ref{eq:FreeLindblad}) with $(\Gamma^{(l)}_1, \Gamma^{(l)}_2, \Gamma^{(r)}_1, \Gamma^{(r)}_2) = (1, 0.6, 1, 0.3)$.}
\label{fig:freeeth}
\end{figure*}
\section{Restricted NESS-ETH in quadratic systems}\label{app:restrictedeth}
In this section, we show the existence of a restricted NESS-ETH in the eigenstates of $\rhoss$ of quadratic systems, and the construction of representative pure states for certain operators.
We work with the XY model with a magnetic field 
\begin{equation}
    H_{XY} = \sumal{j = 1}{N-1}{\left(\frac{1 + \gamma}{2}\sigma^x_j \sigma^x_{j+1} + \frac{1 - \gamma}{2}\sigma^y_j \sigma^y_{j+1}\right)} + \sumal{j = 1}{N}{h \sigma^z_j},
\label{eq:XYmodel}
\end{equation}
with the jump operators
\begin{eqnarray}
    &&L_1 = \sqrt{\Gamma^{(l)}_1}\sigma^+_1 \;\;\; L_2 = \sqrt{\Gamma^{(l)}_2}\sigma^{-}_1 \nn \\
    &&L_3 = \sqrt{\Gamma^{(r)}_1}\sigma^+_N \;\;\; L_4 = \sqrt{\Gamma^{(r)}_2}\sigma^-_N.
\label{eq:FreeLindblad}
\end{eqnarray}
After a Jordan-Wigner transformation, in the Majorana language, the XY Hamiltonian reads\cite{prosen2008third}
\begin{eqnarray}
    &H_{XY} = -i \sumal{j = 1}{N-1}{\left(\frac{1 + \gamma}{2} w_{2j} w_{2j + 1} - \frac{1 - \gamma}{2} w_{2j -1} w_{2j + 1}\right)}\nn \\
    &- i \sumal{j = 1}{N}{ h w_{2j -1} w_{2j}},
\end{eqnarray}
and the corresponding jump operators are equivalent to\cite{prosen2008third}
\begin{eqnarray}
    &&L_1 = \frac{\sqrt{\Gamma^{(l)}_1}}{2}\left(w_1 - i w_2\right),\; L_2 = \frac{\sqrt{\Gamma^{(l)}_2}}{2}\left(w_1 + i w_2\right),\nn \\
    &&L_3 = \frac{\sqrt{\Gamma^{(r)}_1}}{2}\left(w_{2N-1} - i w_{2N}\right),\; L_4 = \frac{\sqrt{\Gamma^{(r)}_2}}{2}\left(w_{2N-1} + i w_{2N}\right).\nn \\
\end{eqnarray}
As described in Sec.~\ref{app:prosen}, $\hamilss$ is quadratic in the Majorana operators and can be obtained in terms of a  ``pairing function" $G_{ij}$ (see Eqs.~(\ref{eq:hamilssquad}) and (\ref{eq:nesshamil})). 
The eigenstates of $\rhoss$ are then the many-body eigenstates of $\hamilss$. 
For quadratic Hamiltonians, a restricted version of ETH is expected to hold in many-body states obtained by occupying the single-particle levels according to the Fermi-Dirac distribution.\cite{khemani2014eigenstate} 
That is, the probability $p_k(\beta)$ of occupying a single-particle level at energy $\mE_k$ is 
\begin{equation}
    p_k(\beta) = \frac{1}{e^{\beta \mE_k} + 1}.
\label{eq:FDdistribution}
\end{equation}
Indeed, as shown in Fig.~\ref{fig:freeeth}, we find that in the many-body eigenstate $\ket{\psi\left(\beta\right)}$ of $\hamilss$, 
\begin{equation}
    \bra{\psi\left(\beta\right)} O \ket{\psi\left(\beta\right)} = f(\beta) + \mathcal{O}\left(\frac{1}{\sqrt{N}}\right),
\label{eq:ETHobservation}
\end{equation}
where $f(\beta)$ is a smooth function of $\beta$, and $O$ is a few-body operator. %
For the construction of representative pure states, we require for a bipartition of the system $A$ and $B$ with $N_A$ and $N_B$ spins such that $N_B \ll N_A$, we require
\begin{equation}
    \textrm{Tr}_B\left(\ket{\psi\left(\beta\right)} \bra{\psi\left(\beta\right)}\right) \approx \textrm{Tr}_B\left(\rhoss\right).
\label{eq:repstatefreeferm}
\end{equation}
Using the fact that $\rhoss$ is a Gibbs state of $\hamilss$ with $\beta = 1$, and $\ket{\psi\left(\beta\right)}$ are the many-body eigenstates of $\hamilss$, it is easy to see that Eq.~(\ref{eq:repstatefreeferm}) is true only if $\beta \approx 1$.\cite{peschel2009reduced}
Thus, as long as the standard deviation of operator expectation values decay with $N$ at $\beta = 1$, one should be able to find a representative pure state. 
However, note that if an operator $\hat{O}_n$ is the occupation number of a single-particle eigenstate (of energy $\mE_n$) of $\hamilss$, then $\bra{\psi\left(\beta\right)} \hat{O} \ket{\psi\left(\beta\right)} = 0, 1$ whereas $\textrm{Tr}\left(\rhoss \hat{O}\right) = 1/(e^{\mE_n} + 1)$, and thus the fluctuations are always $\mathcal{O}(1)$.\cite{khemani2014eigenstate} 
Nevertheless, for operators that are spread over various occupation number operators (for example operators local in real-space), one can expect to find a representative pure state. 
In Fig.~\ref{fig:freeeth}, we plot $f(\beta)$ for various operators and show the corresponding standard deviations in the inset for various values of $\beta$.
The standard deviations decay as $1/\sqrt{N}$ at all temperatures (which correspond to different parts of the many-body spectrum), typical of integrable systems.\cite{alba2015eigenstate}
Note that an interesting aspect of the XY model is the existence of a phase where $\rhoss$ carries long-range correlations.\cite{prosen2008quantum}
As shown in Fig.~4 of Ref.~[\onlinecite{prosen2008quantum}], correlations are long-ranged when $0 < h < 1 - \gamma^2$, and short-ranged otherwise. 
In the long-range phase ($0 < h < 1 - \gamma^2$), the pairing function $G_{ij}$ of Eq.~(\ref{eq:nesspairing}) is no longer local.
That is, $\lim_{|i - j| \rightarrow \infty}{G_{ij}} \neq 0$.
As a consequence of Eqs.~(\ref{eq:hamilssquad}) and (\ref{eq:nesshamil}), $\hamilss$ is a non-local Hamiltonian in the long-range phase.
Nevertheless, as shown in Fig.~\ref{fig:freeeth}b, we find that the locality of $\hamilss$ is irrelevant to the restricted NESS-ETH of $\rhoss$ and to the construction of representative states, although we observe larger finite size effects. 
\bibliography{eth_lindblad}
\end{document}